\documentclass[]{aastex631}

\begin{document}

\title{Cross-Survey Image Transformation: Enhancing SDSS and DECaLS Images to Near-HSC Quality for Advanced Astronomical Analysis}

\correspondingauthor{Zhijian Luo, Jianzhen Chen}
\email{zjluo@shnu.edu.cn, jzchen@shnu.edu.cn}

\author[0009-0009-1617-8747]{Zhijian Luo}
\affiliation{Shanghai Key Lab for Astrophysics, Shanghai Normal University, Shanghai 200234, China}

\author[0000-0001-8485-2814]{Shaohua Zhang}
\affiliation{Shanghai Key Lab for Astrophysics, Shanghai Normal University, Shanghai 200234, China}

\author{Jianzhen Chen}
\affiliation{Shanghai Key Lab for Astrophysics, Shanghai Normal University, Shanghai 200234, China}

\author[0000-0002-2326-0476]{Zhu Chen}
\affiliation{Shanghai Key Lab for Astrophysics, Shanghai Normal University, Shanghai 200234, China}

\author[0000-0003-0688-8445]{Liping Fu}
\affiliation{Shanghai Key Lab for Astrophysics, Shanghai Normal University, Shanghai 200234, China}

\author[0000-0001-8244-1229]{Hubing Xiao}
\affiliation{Shanghai Key Lab for Astrophysics, Shanghai Normal University, Shanghai 200234, China}

\author[0000-0001-9781-6863]{Wei Du}
\affiliation{Shanghai Key Lab for Astrophysics, Shanghai Normal University, Shanghai 200234, China}

\author{Chenggang Shu}
\affiliation{Shanghai Key Lab for Astrophysics, Shanghai Normal University, Shanghai 200234, China}

%% Note that the \and command from previous versions of AASTeX is now
%% depreciated in this version as it is no longer necessary. AASTeX 
%% automatically takes care of all commas and "and"s between authors names.

%% AASTeX 6.31 has the new \collaboration and \nocollaboration commands to
%% provide the collaboration status of a group of authors. These commands 
%% can be used either before or after the list of corresponding authors. The
%% argument for \collaboration is the collaboration identifier. Authors are
%% encouraged to surround collaboration identifiers with ()s. The 
%% \nocollaboration command takes no argument and exists to indicate that
%% the nearby authors are not part of surrounding collaborations.

%% Mark off the abstract in the ``abstract'' environment. 
\begin{abstract}

This study focuses on transforming galaxy images between astronomical surveys, specifically enhancing images from the Sloan Digital Sky Survey (SDSS) and the Dark Energy Camera Legacy Survey (DECaLS) to achieve quality comparable to the Hyper Suprime-Cam survey (HSC). We proposed a hybrid model called Pix2WGAN, which integrates the pix2pix framework with the Wasserstein Generative Adversarial Network with Gradient Penalty (WGAN-GP) to convert low-quality observational images into high-quality counterparts. Our model successfully transformed DECaLS images into pseudo-HSC images, yielding impressive results and significantly enhancing the identification of complex structures, such as galaxy spiral arms and tidal tails, which may have been overlooked in the original DECaLS images. Moreover, Pix2WGAN effectively addresses issues like artifacts, noise, and blurriness in both source and target images. In addition to the basic Pix2WGAN model, we further developed an advanced architecture called Cascaded Pix2WGAN, which incorporates a multi-stage training mechanism designed to bridge the quality gap between SDSS and HSC images, demonstrating similarly promising outcomes. We systematically assessed the similarity between the model-generated pseudo-HSC images and actual HSC images using various metrics, including Root Mean Squared Error (RMSE), Peak Signal-to-Noise Ratio (PSNR), and Structural Similarity Index (SSIM), along with perceptual metrics such as Learned Perceptual Image Patch Similarity (LPIPS) and Fréchet Inception Distance (FID). The results indicate that images transformed by our model outperform both the original SDSS and DECaLS images across nearly all evaluation metrics. Our research is expected to provide significant technical support for astronomical data analysis, cross-survey image integration, and high-precision astrometry.

\end{abstract}

%% Keywords should appear after the \end{abstract} command. 
%% The AAS Journals now uses Unified Astronomy Thesaurus concepts:
%% https://astrothesaurus.org
%% You will be asked to selected these concepts during the submission process
%% but this old "keyword" functionality is maintained in case authors want
%% to include these concepts in their preprints.
\keywords{Galaxies (573) --- Convolutional neural networks
(1938) --- Ground-based astronomy (686) ---  Astronomy data modeling (1859) --- Astronomy
data analysis (1858) --- Computational astronomy (293) --- Astronomy data
visualization (1968)}

%% From the front matter, we move on to the body of the paper.
%% Sections are demarcated by \section and \subsection, respectively.
%% Observe the use of the LaTeX \label
%% command after the \subsection to give a symbolic KEY to the
%% subsection for cross-referencing in a \ref command.
%% You can use LaTeX's \ref and \label commands to keep track of
%% cross-references to sections, equations, tables, and figures.
%% That way, if you change the order of any elements, LaTeX will
%% automatically renumber them.
%%
%% We recommend that authors also use the natbib \citep
%% and \citet commands to identify citations.  The citations are
%% tied to the reference list via symbolic KEYs. The KEY corresponds
%% to the KEY in the \bibitem in the reference list below. 

\section{Introduction} \label{sec:intro}

In astronomical research, galaxy images are essential resources that provide critical visual evidence of galaxy structures and evolution, as well as vital data for understanding the formation and development of the universe. For example, the citizen science project Galaxy Zoo engages the public in classifying and analyzing millions of galaxy images, assisting scientists in further investigations into galaxy formation and evolution, the structure of the universe, and the effects of dark matter and dark energy \citep{lintott2008galaxy,lintott2011galaxy,willett2013galaxy}.

As exploration of galaxies and the universe progresses, research increasingly relies on high-quality observational images. These high-quality images can reveal detailed structural features of galaxies, such as spiral arms, bars, and the distribution of interstellar gas and dust, as well as the dynamics of active star-forming regions. This information is crucial for studying the evolutionary history of galaxies, the physical state of the interstellar medium, and the environments in which star formation occurs \citep{madau2014cosmic}. Moreover, high-quality galaxy images facilitate the examination of galaxy interactions, including mergers, gravitational interactions, and star formation activities. These phenomena not only illuminate the lifecycle of galaxies but also enhance our understanding of their roles within the larger cosmic structure and their impacts on cosmic evolution \citep{conselice2003relationship,conselice2014evolution}.

Furthermore, high-quality images of galaxies play a critical role in the study of dark matter and large-scale cosmic structures. Through precise imaging, scientists can utilize gravitational lensing to infer the distribution of dark matter, leading to a deeper understanding of its influence on galaxy formation and evolution \citep{massey2010dark}. Thus, the quality of observational images directly influences the extent of scientific discovery.

To improve image quality, the most straightforward method is to enhance the hardware of astronomical telescopes, commonly referred to as the “hardware approach”. Common hardware improvements include: 1) increasing the telescope aperture to enhance resolution and light-gathering capability \citep{roggemann1997improving,early2004twenty,cai2019toward}; 2) equipping telescopes with adaptive optics systems to compensate for atmospheric turbulence \citep{glindemann2000adaptive,hickson2014atmospheric,hippler2019adaptive}; 3) advancing photodetector technology to reduce noise and improve detection efficiency \citep{bai2008teledyne,lesser2015summary,kitchin2020astrophysical}; and 4) optimizing the design of optical components to minimize aberrations and increase light transmission \citep{paez2001telescopes,breckinridge2015polarization}. While these hardware upgrades can significantly enhance the quality of astronomical images, they face challenges such as high manufacturing costs and material strength limitations, meaning that telescope performance cannot be improved indefinitely. Additionally, external factors such as atmospheric conditions, background noise, and light pollution further restrict the effectiveness of hardware improvements during observations \citep{popowicz2016efficiency,schawinski2017generative,jia2021data}.

Due to the limitations of hardware enhancements, “software methods” have become another important means of improving image quality. Traditional software techniques include image deconvolution, background noise removal, deblurring, and image stacking. For instance, wavelet transforms and deconvolution algorithms can be utilized to reduce noise and restore image details \citep{magain1998deconvolution,courbin1999deconvolving,starck2002deconvolution,wang2023neural,sreejith2024point}. Additionally, flat field correction and dark current correction techniques can remove background noise, enhancing the contrast of celestial objects \citep{blanton2011improved,ma2014new}. Stacking multiple observational datasets can also improve the signal-to-noise ratio and increase the visibility of faint celestial objects \citep{zibetti2007optical,morales2018systematic}. However, these methods are constrained by the Shannon-Nyquist sampling theorem \citep{nyquist1928certain,shannon1949communication}. In situations with high noise and low-resolution images, traditional techniques may struggle to adequately restore details, thereby affecting the accurate study of celestial objects \citep{magain1998deconvolution,magain2007deconvolution,cantale2016firedec}.

Recently, rapid advancements in machine learning and artificial intelligence have led to data-driven image enhancement techniques across various fields. These technologies employ intelligent algorithms to recover image details in complex backgrounds and high-noise conditions, continually pushing the boundaries of image processing and often surpassing the diffraction limits of optical systems. They are widely used in medical imaging, autonomous driving, remote sensing, and other areas, significantly improving image analysis and decision-making capabilities.

However, the use of image enhancement techniques in astronomy remains relatively limited compared to other fields. This limitation is primarily due to the unique characteristics of astronomical images, such as high noise levels, low signal-to-noise ratios, and complex imaging conditions, which present additional processing challenges. Nevertheless, as research advances, more astronomers are exploring machine learning and image enhancement technologies to improve image quality and increase analytical accuracy. These applications encompass various areas, including denoising \citep{vojtekova2021learning,sweere2022deep}, deblending \citep{boucaud2020photometry,arcelin2021deblending,wang2022galaxy},deblurring \citep{reiman2019deblending}, image restoration \citep{schawinski2017generative,long2021learning}, and super-resolution reconstruction \citep{li2018super,miao2024astrosr}.

For instance, \citet{dou2022super} utilized a deep learning model called DownSampleGAN (DSGAN) to perform super-resolution reconstruction of magnetic field maps from the Michelson Doppler Imager (MDI), successfully achieving spatial resolution comparable to that of images from the Helioseismic and Magnetic Imager (HMI). \citet{yang2023image} applied three typical deep learning super-resolution models to enhance the resolution of FY-3E/X-EUVI 195 Å images, aligning them with the spatial resolution of Atmospheric Imaging Assembly (AIA) 193 Å images from the Solar Dynamics Observatory. Additionally, this study improved the temporal resolution of full-disk EUV solar images.

Notably, \citet{schawinski2017generative} proposed the GalaxyGAN model based on conditional Generative Adversarial Networks (GANs), which was trained using samples from SDSS galaxy images to successfully recover high-detail original features from artificially degraded low-quality galaxy images. Since then, more advanced artificial intelligence algorithms have emerged to enhance original images and generate higher-resolution outputs, including the Enhanced Deep Super-Resolution Network (EDSR; \citealt{lim2017enhanced}), the Very Deep Residual Channel Attention Network (RCAN; \citealt{zhang2018image}), the Efficient Non-Local Contrastive Network (ENLCN; \citealt{xia2022efficient}), and the Super-Resolution Generative Adversarial Network (SRGAN; \citealt{ledig2017photo}). These algorithms have been widely applied in the super-resolution reconstruction tasks of astronomical images \citep{li2022self,shibuya2024galaxy,miao2024astrosr}, demonstrating the tremendous potential of artificial intelligence algorithms in reconstructing details of galaxy images, thus maximizing resource utility and reducing the costs and time of repeated observations, which enhances the interpretability and scientific value of astronomical data \citep{kinakh2024hubble}.

However, most current algorithms primarily focus on improving the resolution of observational images, while resolution enhancement is only one aspect of overall image quality improvement. In reality, enhancing image quality also involves several key factors, such as contrast, color accuracy, detail retention, and noise suppression. Therefore, solely increasing the resolution while neglecting these other factors may limit the perceived quality of the generated images and their usefulness in scientific research.

Currently, another approach that is gaining attention among researchers is the use of deep learning techniques to transform images between different telescope observations, thereby improving the overall quality of existing low-quality observational images \citep{buncher2021survey2survey}. This serves as a proof of concept for the potential of such methodologies in astronomical image processing. However, due to variations in observational conditions, the complexity of image features, and challenges in algorithm adaptability, research in this area remains relatively limited. Therefore, there is an urgent need to develop new effective methods for mapping and transforming images from diverse observational facilities.

In this study, we present a novel hybrid model called Pix2WGAN, which builds upon the classical pix2pix framework \citep{isola2017image}. Our model combines the strengths of pix2pix with the Wasserstein GAN with Gradient Penalty (WGAN-GP) \citep{adler2018banach}. The original pix2pix model has been widely applied in various fields such as image-to-image translation, image restoration, style transfer, and satellite image analysis, demonstrating exceptional performance and flexibility across these tasks. By integrating WGAN-GP, our model enhances both the quality and stability of the generated images. Recent research has increasingly focused on the potential of such hybrid models in various computer vision applications; for instance, \citet{tirel2024novel} successfully employed a similar model for denoising text images.

The primary aim of our work is to facilitate data transformation between different astronomical observation projects while enhancing the overall quality and detail representation of low-quality astronomical images. This work also serves as a proof of concept for using deep learning techniques in the transformation of astronomical data. We apply the Pix2WGAN model to galaxy image transformation across three major astronomical survey datasets: the Sloan Digital Sky Survey (SDSS), the Dark Energy Camera Legacy Survey (DECaLS), and the Hyper Suprime-Cam Survey (HSC).

We note that \citet{miao2024astrosr} investigated the transformation of SDSS to HSC images using four GAN-based super-resolution models \citep{ledig2017photo, lim2017enhanced, zhang2018image, xia2022efficient}. However, their emphasis on integer scaling factors required downsampling HSC images to match the resolution of SDSS, which could compromise the quality of the generated pseudo-HSC images. In contrast, our Pix2WGAN does not rely on maintaining an integer scaling factor, allowing for the direct generation of high-quality pseudo-HSC galaxy images from low-quality SDSS or DECaLS data, thereby avoiding the quality loss associated with downsampling.

%To evaluate the quality of the images generated by our model, we use both quantitative and qualitative assessment methods to compare the generated pseudo-HSC images with their corresponding observed HSC images. This evaluation not only verifies the reliability of the generated images but also offers a comprehensive analysis of the model’s performance in detail recovery and overall quality enhancement.

Our model can generate high-quality pseudo-HSC datasets within the observational areas of SDSS and DECaLS, as well as in regions that HSC has not yet observed. Additionally, the scalability of the Pix2WGAN model enhances its applicability for image transformation across various survey projects, enabling large-scale astronomical data processing across platforms and the production of multi-band images, thereby addressing future requirements for astronomical data integration.

The structure of this paper is organized as follows: Section 2 provides a brief overview of the data from the three observational projects used in this study—SDSS, DECaLS, and HSC—and outlines the data collection and sample construction processes. Section 3 provides a detailed analysis of the architecture and training process of the Pix2WGAN model. In Section 4, we present the application results of this model on SDSS, DECaLS, and HSC images, along with a comprehensive quality assessment of the generated pseudo-HSC images using both quantitative and qualitative evaluation methods. Finally, Section 5 summarizes the research findings and discusses the implications of our results.

\section{Data Overview and Collection} \label{section:Data}

The data used in this work are sourced from three survey projects: SDSS, DECaLS, and HSC. The differences in spatial resolution and depth among these surveys result in significant variations in their galaxy images. Here, we provide a brief overview of the imaging data from these three surveys, followed by an explanation of our sample collection process.

\subsection{SDSS Images} \label{sec:sdss}

The Sloan Digital Sky Survey (SDSS) \footnote{\url{https://www.sdss.org/}} is a large-scale astronomical survey project that began in 2000 \citep{york2000sloan}. It has produced the most detailed three-dimensional map of the universe to date, covering approximately one-third of the sky. SDSS provides multi-band deep images ($u$, $g$, $r$, $i$, $z$ bands) and spectral data for over 3 million celestial objects, significantly advancing our understanding of galaxies, dark matter, and large-scale structures in the universe \citep{gunn20062,abazajian2009seventh}.

The SDSS project employs a 2.5-meter telescope located at Apache Point Observatory in New Mexico, USA, which is well-suited for extensive deep-sky observations. However, the telescope’s imaging depth and resolution (typically ranging from 1.5 to 2 arcseconds) are relatively low compared to subsequent high-resolution surveys, limiting its capability to capture fine details. The charge-coupled device (CCD) sensors used in the SDSS camera have a pixel resolution of 0.396 arcseconds per pixel (in the $r$ band). Despite processing techniques such as noise reduction and background light correction, these data may still struggle to resolve the intricate structures of galaxies.

While SDSS data is extensive and provides a wealth of astronomical information, its limitations in spatial resolution and depth hinder detailed structural studies compared to higher-resolution projects like DECaLS and HSC.

\subsection{DECaLS Images} \label{sec:decals}

The Dark Energy Camera Legacy Survey (DECaLS) \footnote{\url{https://www.legacysurvey.org/decamls/}} employs the Dark Energy Camera (DECam; \citealt{flaugher2015dark}) to observe cosmic phenomena such as dark matter, dark energy, and galaxy evolution. DECam is mounted on the 4-meter Blanco telescope in Chile and features a wide field of view of 3.2 square degrees, with a pixel resolution of 0.262 arcseconds per pixel. Compared to SDSS, DECaLS significantly enhances image resolution, achieving full widths at half maximum (FWHM) of $1.29^{\prime\prime}$, $1.18^{\prime\prime}$, and $1.11^{\prime\prime}$ in the $g$, $r$, and $z$ bands, respectively. Overall, the image resolution ranges from 0.6 to 1 arcsecond, enabling the detection of finer structures in celestial objects \citep{depoy2008dark}. 

The DECaLS survey plan provides approximately two-thirds of the optical imaging coverage for the ongoing Dark Energy Spectroscopic Instrument (DESI; \citealt{dey2019overview}), focusing primarily on the northern Galactic cap (declination $\leq 32$ degrees) and the southern Galactic cap (declination $\leq 34$ degrees). Although the coverage area is extensive, there remains potential for further expansion compared to SDSS. In this study, we use the most up-to-date data release (DR15).

\subsection{HSC Images} \label{sec:hsc}

The Hyper Suprime-Cam Subaru Strategic Survey (HSC-SSP) \footnote{\url{https://hsc-release.mtk.nao.ac.jp/doc/ }} is another significant astronomical project focused on studying topics such as galaxy evolution, gravitational lensing, supernovae, and galactic structure \citep{miyazaki2012mclean,miyazaki2018hyper}. HSC is mounted on the 8.2-meter Subaru telescope in Japan and features a field of view with a diameter of 1.5 degrees, as well as a pixel resolution of 0.168 arcseconds per pixel . The FWHM in the $i$ band is 0.6 arcseconds, enabling it to resolve the structures of spiral galaxies and other celestial details.

The HSC-SSP survey is divided into three components: wide, deep, and ultra-deep fields, which overlap with those of the SDSS \citep{ahumada202016th}, the Baryon Oscillation Spectroscopic Survey (BOSS; \citealt{dawson2012baryon}), the Galaxy and Mass Assembly survey (GAMA; \citealt{driver2011galaxy}) and the Visible Multi-Object Spectrograph Very Large Telescope Deep Survey (VVDS; \citealt{le2005vimos}). The third data release (PDR3), published in 2021, covers approximately 670 square degrees across five filters ($g, r, i, z, y$) and achieves a depth of 26 magnitude ($5\sigma$). While the image quality and resolution of HSC surpass those of SDSS and DECaLS, its coverage area remains relatively small.

\subsection{Data Collection} \label{sec:data_collection}

The main objective of this study is to utilize advanced deep learning methods to convert low-quality images from SDSS and DECaLS into high-quality images comparable to those from HSC. To achieve this, we need to construct a large sample that includes images from these survey projects. Our sample is derived from galaxies in the Galaxy Zoo DECaLS project, all of which are spectroscopic targets from SDSS. Consequently, the sample primarily includes galaxies with brightness greater than $m_r = 17.77$ and covers redshifts of $z = 0.15$ or lower. Additionally, these galaxies are required to have a Petrosian radius (as specified in the PETROTHETA4 column of the NASA–Sloan Atlas v1.0.0) of at least 3 arcseconds \citep{walmsley2022galaxy}.

We first cross-matched the overlapping regions of these three observational projects and randomly selected 15,000 galaxies from these areas as training samples. This selection ensures that the model we build can learn a diverse range of features, including different types and morphologies of galaxies. Furthermore, we selected an additional 2,000 galaxies from these overlapping regions as test samples to evaluate the model’s generalization capabilities and the quality of the generated images. All images of these galaxies were obtained from the Legacy Survey website \footnote{\url{https://www.legacysurvey.org/}}.

The Legacy Survey website serves as a comprehensive platform for an astronomical project designed to showcase and provide access to observational data produced by DESI Legacy imaging survey. This website consolidates image data from various astronomical projects, including SDSS, DECaLS, HSC, and the Wide-field Infrared Survey Explorer (WISE). Users can access astronomical images and databases of celestial objects generated by the Legacy Survey, allowing them to browse imaging and photometric information for billions of celestial bodies. Additionally, the website provides various interactive tools to help users acquire and analyze this data, which is significant for research on dark energy, galaxy and star evolution, and the exploration of large-scale structures in the universe.

We used the JPEG cropping tool \footnote{\url{https://www.legacysurvey.org/viewer/}} provided by the Legacy Survey website to download the corresponding images from SDSS, DECaLS, and HSC based on the right ascension and declination coordinates of the sample galaxies. To ensure consistency in the spatial coverage of images for the same galaxy, we utilized the algorithms provided by the website to adjust the pixel scale of all images to match that of HSC, which has a pixel scale of 0.168 arcseconds per pixel.

The Legacy Survey website also offers various band combinations for color synthesis of the images. After visual inspection, we selected the $g$, $r$, and $i$ bands for the synthesis of SDSS and HSC images, while for DECaLS images, we chose the $g$, $r$, and $z$ bands due to the survey’s primary focus on observations in these bands. This selection ensures that the final generated images maintain visually consistent colors, making images of galaxies from different data sources appear more harmonious and facilitating subsequent scientific analysis and comparison.

Due to hardware resource limitations, we constrained the image size to $128 \times 128$ pixels. This size is sufficient to capture enough background sky while preserving the key structures of most galaxies, ensuring that the model can process efficiently under limited resource conditions. It is important to note that our model does not have a strict dependency on the size of the input images; larger or smaller sized images are also applicable. With appropriate adjustments, the model can maintain relatively stable performance across images of different sizes.

\begin{figure} 
        \includegraphics[width=\textwidth]{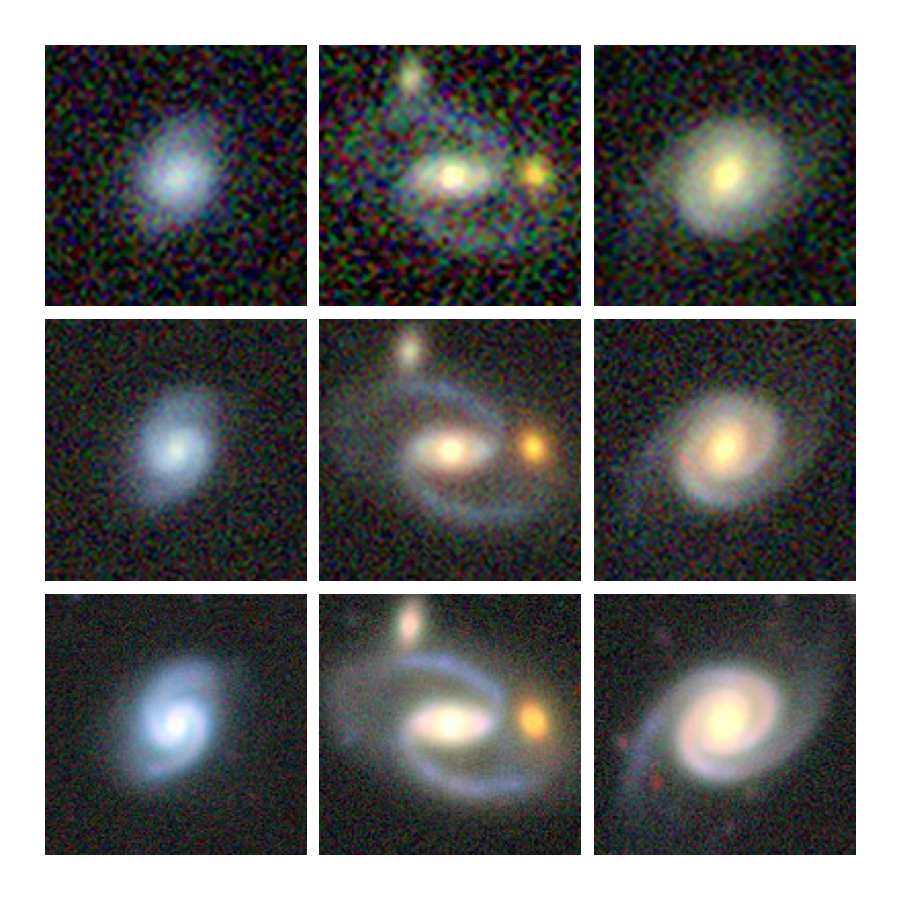} 
        \caption{Example images of three groups of galaxies. The panels, arranged from left to right, showcase images of different galaxies, while those arranged from bottom to top correspond to the observations of the same galaxy from three surveys with varying depths and resolutions: HSC, DECaLS, and SDSS. The images used are sourced from publicly available data on the official Legacy Survey website.} 
        \label{fig:sdss_decals_hsc} 
\end{figure}

Figure \ref{fig:sdss_decals_hsc} presents three examples of sample galaxies, displaying the galaxy images observed in HSC, DECaLS, and SDSS from bottom to top. It is clear that with increased exposure depth and resolution, the structures of the galaxies become more pronounced, and their sizes appear more extended. In images with shallower exposure depths, such as those from the SDSS project, fine structures in the outer regions of the galaxies are difficult to discern. However, with the greater exposure depth observed in the HSC project, these structures become clearer, revealing richer and more distinct internal details of the galaxies.

Our final sample consists of a training set made up of 15,000 groups of galaxies, while the test set consists of 2,000 groups of galaxies, with each group containing the corresponding three images from SDSS, DECaLS, and HSC. All images are adjusted to a resolution of 0.168 arcseconds per pixel and sized at $128 \times 128$ pixels, with pixel values ranging from [0, 255], representing the intensity of the colors in the images.

\section{Methodology}

In this section, we will first provide a detailed introduction to the classic pix2pix model, which is widely used across various fields, primarily for image-to-image translation. Next, we will explain our improvements to this model by integrating WGAN-GP into the pix2pix framework, thereby constructing the Pix2WGAN hybrid model. To the best of our knowledge, this paper represents the first application of this hybrid model in the field of astronomy, with the aim of transforming SDSS and DECaLS images into high-quality images comparable to those from the HSC survey project.

\subsection{Classical pix2pix Model}

Pix2pix is an image translation model based on conditional GAN, proposed by \citet{isola2017image} at the 2017 CVPR (IEEE Conference on Computer Vision and Pattern Recognition). It is widely used for tasks that involve transforming one image style into another, including converting low-resolution images to high-resolution images, transforming sketches into realistic images, and changing daytime scenes to nighttime scenes \citep{chen2018sketchygan,chao2019high,liu2020sketch,henry2021pix2pix,liu2022night,kumar2024artistic}.

\begin{figure}
	\includegraphics[width=\textwidth]{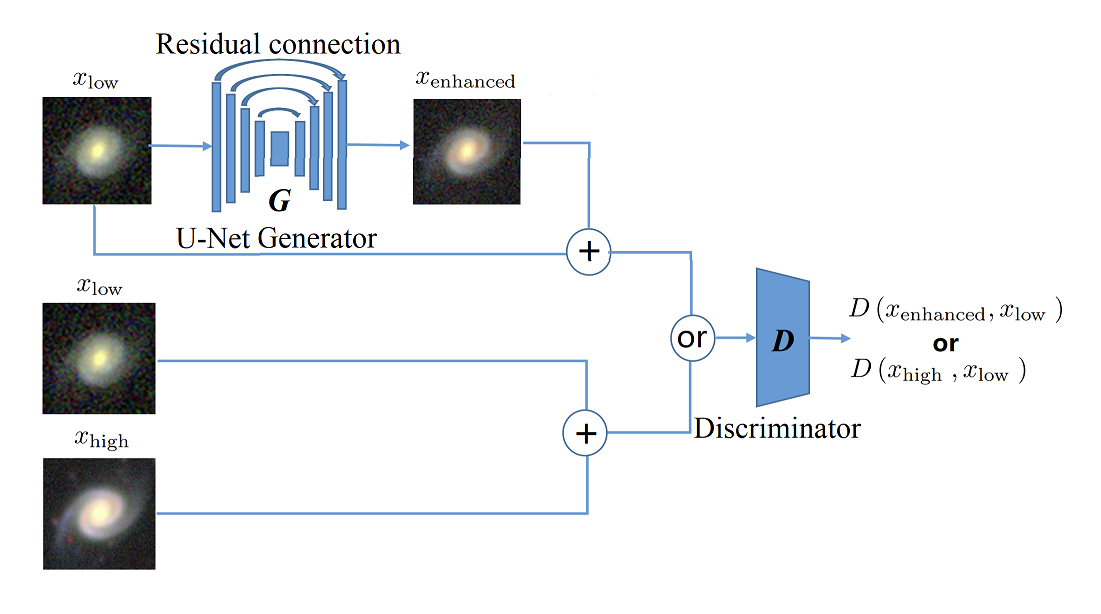}
	\caption{Schematic diagram of the classical pix2pix network architecture. $x_{\text{low}}$ represents the low-quality input image, $x_{\text{enhanced}}$ represents the enhanced image generated by the generator after transformation, and $x_{\text{high}}$ represents the high-quality target/real image. $G$ is the generator, which adopts a U-Net structure, while $D$ is the discriminator, whose task is to distinguish between the authenticity of the input image pairs.}
	\label{fig:pix2pix_archi}
\end{figure}

The core concept of pix2pix is to generate images that meet specific conditions using a conditional GAN. The model learns the mapping relationship between paired input and target images, enabling it to produce output images corresponding to the input images. Its network architecture, as illustrated in Figure \ref{fig:pix2pix_archi}, consists primarily of two components: a generator and a discriminator.

The generator employs a U-Net architecture \citep{ronneberger2015u,isola2017image}, which is a convolutional neural network (CNN) comprising an encoder and a decoder, specifically designed to transform input images into target images. In our specific task, the encoder is designed to consist of six modules, each composed of a convolutional layer, a batch normalization layer, and a Leaky ReLU activation function. This configuration progressively extracts low-level features while downsampling the images to capture essential information.

The decoder is similarly constructed with six modules, where each module includes a transposed convolution layer, a batch normalization layer, dropout layers (applied to the first three modules), and a Leaky ReLU activation function. This design gradually restores the spatial resolution of the images. In the U-Net architecture, the convolutional layers in the encoder are connected to the corresponding layers in the decoder through skip connections. This design effectively combines low-dimensional features with high-resolution features, preserving fine details and resulting in generated images that are more refined and realistic.

The discriminator employs a conditional PatchGAN architecture \citep{demir2018patch}. A key feature of the PatchGAN is its ability to divide input images into multiple patches and evaluate their authenticity through local judgments of each patch. This approach not only emphasizes the consistency of local features but also enables the model to focus on details within each patch, effectively enhancing the quality of detail in the generated images.

In our specific task, the discriminator consists of five convolutional layers, each followed by a batch normalization layer and a Leaky ReLU activation function, ultimately producing a patch with the shape of (batchsize, 14, 14, 1). This patch is then averaged along the width and height dimensions to obtain a scalar value, which serves as the final output of the discriminator. The discriminator receives two sets of image pairs as input: one set contains the images outputted by the generator along with their corresponding input images, while the other set includes real/target images and their respective input images, with each set of image pairs concatenated along the channel dimension. The primary task of the discriminator is to learn to differentiate between the authenticity of these two sets of image pairs.

During the training process, the classical pix2pix model optimizes image transformation results primarily through an adversarial learning mechanism between the generator and the discriminator. The model's total cost function is divided into generator loss and discriminator loss, with the generator loss comprising both adversarial loss and structural loss. 

The total loss for the classic pix2pix generator can be expressed as:
\begin{equation}
    J_{1, \text {gen}}(G, D)=J_{1, \text {adv}}(G, D)+\lambda * J_{1, \text {str}}(G),
\end{equation}
where $G$ denotes the generator and $D$ denotes the discriminator. $J_{1, \text {adv}}(G, D)$ represents the adversarial loss, which is calculated using Binary Cross Entropy (BCE) \footnote{\url{https://www.tensorflow.org/api_docs/python/tf/keras/losses/BinaryCrossentropy}} to maximize the probability that the discriminator recognizes the generated image as a real image: $B C E\left(D\left(x_{\text {enhanced}}, x_{\text {low }}\right), y_{\text {true }}\right)$, where $x_{\text{low}}$ represents the original low-quality input image, $x_{\text{enhanced}}$ refers to the enhanced image generated by the generator, and $y_{\text{true}}$ is a vector of ones of the same size as the batch. $J_{1, \text {str}}(G)$ represents the structural loss, which uses $L1$ loss to quantify the difference between the generated image and the target image: $L1\left(x_{\text {enhanced}}, x_{\text {high }}\right)$, ensuring that the generated image retains detailed information, where,$x_{\text {high}}$ refers to the high-quality target image. $\lambda$ is a hyperparameter that balances the adversarial loss and the structural loss, and its value is typically tuned according to the specific task at hand. 

The discriminator loss evaluates the discriminator’s ability to distinguish between the target image and the generated image, with the calculation formula expressed as follows: 
\begin{equation}
    J_{1, \text{dis}}(G, D) = \frac{1}{2} \left( B C E\left(D\left(x_{\text{high}}, x_{\text{low}}\right), y_{\text{true}}\right) + B C E\left(D\left(x_{\text{enhanced}}, x_{\text{low}}\right), y_{\text{false}}\right) \right) 
	\label{eq:loss_str}
\end{equation}
where $y_{\text{false}}$ is a vector of zeros of the size of the batch.

In each iteration of the training process, the generator first generates an image and calculates its loss. It then updates its weights using backpropagation with the goal of minimizing the total loss of the generator. Meanwhile, the discriminator’s loss is based on its evaluation of both real and generated images, aimed at optimizing its ability to distinguish between them. The discriminator updates its weights accordingly to reduce its loss.

Through this alternating update mechanism, the generator and discriminator mutually reinforce each other throughout the training process, continuously enhancing their performance. As training progresses, the generator increasingly produces high-quality images with a strong sense of realism, thereby ensuring the model’s effectiveness and stability in image-to-image translation tasks.

\subsection{Hybrid Pix2pix WGAN-GP Model }

The classic pix2pix model is based on conditional GAN. However, training GANs is often unstable, leading to fluctuations in the quality of the generated images and potentially resulting in mode collapse \citep{thanh2020catastrophic}. Mode collapse refers to a phenomenon where the generator exploits the weaknesses of the discriminator by optimizing along specific paths in the parameter space that can achieve high scores from the discriminator, yet fail to produce meaningful or diverse outputs. As a result, the generator begins to output only a limited range of results, failing to capture the rich features of the training data adequately. To address these issues, we employ WGAN-GP \citep{gulrajani2017improved,chen2021challenges} to improve the loss function structure of the classic pix2pix model and apply it to the specific task at hand . We refer to the modified model as Pix2WGAN.

Specifically, unlike the classical pix2pix model, which utilizes Jensen-Shannon (JS) divergence for binary classification in the discriminator, Pix2WGAN is trained by minimizing the Wasserstein distance between the distributions of real and generated data \citep{adler2018banach}. The Wasserstein distance, also known as the “Earth Mover’s” distance, quantifies the amount of “work” required to transform one probability distribution into another and serves as an effective method for measuring the distance between two distributions. A key advantage of this approach is its ability to provide a smoother and more continuous optimization curve, thereby mitigating the gradient vanishing problem that can arise with JS divergence in certain scenarios.

Additionally, WGAN-GP incorporates a Gradient Penalty technique to ensure that the gradients of the discriminator remain within an appropriate range across all input samples, thus satisfying the 1-Lipschitz condition. This method further enhances the model’s stability, enabling WGAN-GP to effectively generate high-quality samples across a wider range of application scenarios.

In our Pix2WGAN, the total loss function for the generator is:
\begin{equation}
    J_{2, \text {gen}}(G, D)=J_{2, \text {adv}}(G, D)+\lambda * J_{2, \mathrm{str}}(G),
	\label{eq:loss_gen_wgan}
\end{equation}
where, $\lambda$ is the hyperparameter that balances the adversarial loss and structural loss; in our study, it is set to a value of 100. The adversarial loss function of the generator $J_{2, \text {adv}}(G, D)$ is completely different from the classical pix2pix adversarial loss $J_{1, \text {adv}}(G, D)$, and its form has been modified to $-D\left(x_{\text {enhanced }}, x_{\text {low }}\right)$, and the structural loss term $J_{2, \mathrm{str}}(G)$ remains unchanged and is calculated using $L1$ loss as before, i.e., $L 1\left(x_{\text {enhanced }}, x_{\text {high }}\right)$.

For the discriminator, the loss function is also significantly different from that of pix2pix and has been modified to consist of two parts. The first part employs an averaging strategy similar to that of the classical pix2pix model, but its purpose is to measure the difference between real predictions and fake predictions:
\begin{equation}
    J_{2, \text {diff}}(G, D)=D\left(x_{\text {high }}, x_{\text {low }}\right)-D\left(x_{\text {enhanced }}, x_{\text {low }}\right).
	\label{eq:loss_fir_dis_wgan}
\end{equation}

The second part is the gradient penalty term, which is used to enforce gradient penalties. This term is defined by first setting $x_{\text{interp}}$ as the interpolation between the target image and the generated image, as follows:
\begin{equation}
    x_{\text {interp }}=\alpha * x_{\text {high }}+(1-\alpha) * x_{\text {enhanced }},
	\label{eq:loss_interp}
\end{equation}
where $\alpha \in (0, 1)$ is a random number. The gradient penalty is then defined as:
\begin{equation}
    J_{2, \text{gp}}(G, D)=E\left[\left(\left\|\nabla_{x_{\text {interp }}} D\left(x_{\text {interp }}, x_{\text {low }}\right)\right\|_2-1\right)^2\right],
	\label{eq:loss_gp}
\end{equation}
Where $E$ denotes the mathematical expectation. These two terms are combined to define the loss of the discriminator, as follows:
\begin{equation}
   J_{2, \text{dis}}(G, D)=J_{2, \text{diff}}(G, D)+w_p * J_{2, \text{gp}}(G, D),
	\label{eq:loss_wgan_gp}
\end{equation}
where $w_p$ is the weight for the gradient penalty, which is set to 10 in our study.

After improving the loss function of the classical pix2pix model, we developed a new Pix2WGAN hybrid model that exhibits substantial stability during the training process. In the next section, we will present the remarkable performance of this model in the task of transforming SDSS and DECaLS images into HSC-level images.

\section{Experiments and Results} \label{section:experiments}

In this section, we apply the Pix2WGAN model to three major astronomical surveys: SDSS, DECaLS, and HSC, with the goal of transforming low-quality images from SDSS and DECaLS into high-quality images that closely resemble HSC observations. To evaluate the performance of Pix2WGAN in these image transformation tasks, we conduct both qualitative (visual inspection) and quantitative (metric assessment) analyses to compare the pseudo-HSC images generated by the model with the actual HSC images, as well as to assess them against the original images before transformation. Our focus will be on the overall quality, detailed structure, and denoising effectiveness of the transformed galaxy images.

\subsection{Data Preprocessing} \label{subsection:data_prep}
%We start the process by preprocessing both the training and testing samples, a critical step for ensuring the model’s performance. Based on the data preparation outlined in Section 2.4, our training dataset comprises 15,000 galaxies, with each galaxy represented by three RGB images corresponding to low-resolution SDSS, mid-resolution DECaLS, and high-resolution HSC observations. These three sets of images cover the same celestial region and each measures $128 \times 128$ pixels. The testing dataset consists of 2,000 samples, which can be expanded as necessary.

%To enhance the training effectiveness of the model, we performed normalization on the pixel values of the sample images. In the downloaded raw images, all pixel values are clipped to the range of [0, 255]. To accelerate model training and better preserve the detailed information in the images, we further scaled the pixel values of each image to the range of [-1, 1]. Specifically, let the original data be denoted as $x$ ; we calculate the scaled data $x^*$ using the following formula:
Data preprocessing is a critical step in ensuring the model’s performance. Before feeding the images into the model, we normalized the pixel values of the input images. In the downloaded raw images, all pixel values are restricted to the range of [0, 255]. To accelerate the model’s training speed and better retain the details in the images, we further scaled the pixel values of each image to the range of [-1, 1]. Specifically, let the original data be denoted as $x$; we calculate the scaled data $x^*$ using the following formula:

\begin{equation}
     x^* = \frac{x - 127.5}{127.5}.
     \label{eq:normal} 
\end{equation}
In this formula, 127.5 serves as the midpoint of the [0, 255] range. This scaling technique centers the pixel values while maintaining relative contrast, thus aiding in accelerating the model’s convergence and improving the quality of the generated images.

%In this formula, 127.5 serves as the midpoint of the [0, 255] range. This scaling technique centers the pixel values while maintaining relative contrast, thus aiding in accelerating the model’s convergence and improving the quality of the generated images.

\subsection{Transforming Images from DECaLS to HSC} \label{subsection:DECaLS_HSC}

In image-to-image transformation, a smaller difference between images typically simplifies the generator’s task, leading to generated images that more closely match the target images. This reduced disparity helps the model learn the mapping between input and target images more effectively. Given that the difference between DECaLS and HSC images is smaller than that between SDSS and HSC images, this subsection will begin by applying the Pix2WGAN model for the transformation from DECaLS to HSC.

We implemented the Pix2WGAN model using the Keras and TensorFlow 2 libraries \citep{abadi2016tensorflow}, and conducted training and testing on a computer equipped with an NVIDIA RTX 3090 GPU. During training, we set the batch size to 32. Both the generator and discriminator utilized the ADAM optimizer \citep{kingma2014adam}, configured with parameters $\beta_1 = 0.5$ and $\beta_2 = 0.999$. The initial learning rate was set to 0.0002 and was reduced to $1/1.00004$ of its previous value after each epoch. To improve the stability and effectiveness of the training process, we adopted a strategy of updating the generator once every three iterations of training the discriminator. This approach helps maintain a balance between the model’s generative and discriminative capabilities, effectively enhancing the transformation from low-quality to high-quality images. The training was concluded after 500 epochs, with a total training time of approximately 7.3 hours.

To qualitatively evaluate the performance of the Pix2WGAN model and the effects of image transformation, we visually examined the generated images and selected six representative galaxies from the testing set for presentation. These galaxies were chosen based on their diverse morphological features and varying complexities, enabling us to assess the model’s ability to capture different structures and details effectively.

\begin{figure}
	\includegraphics[width=\textwidth]{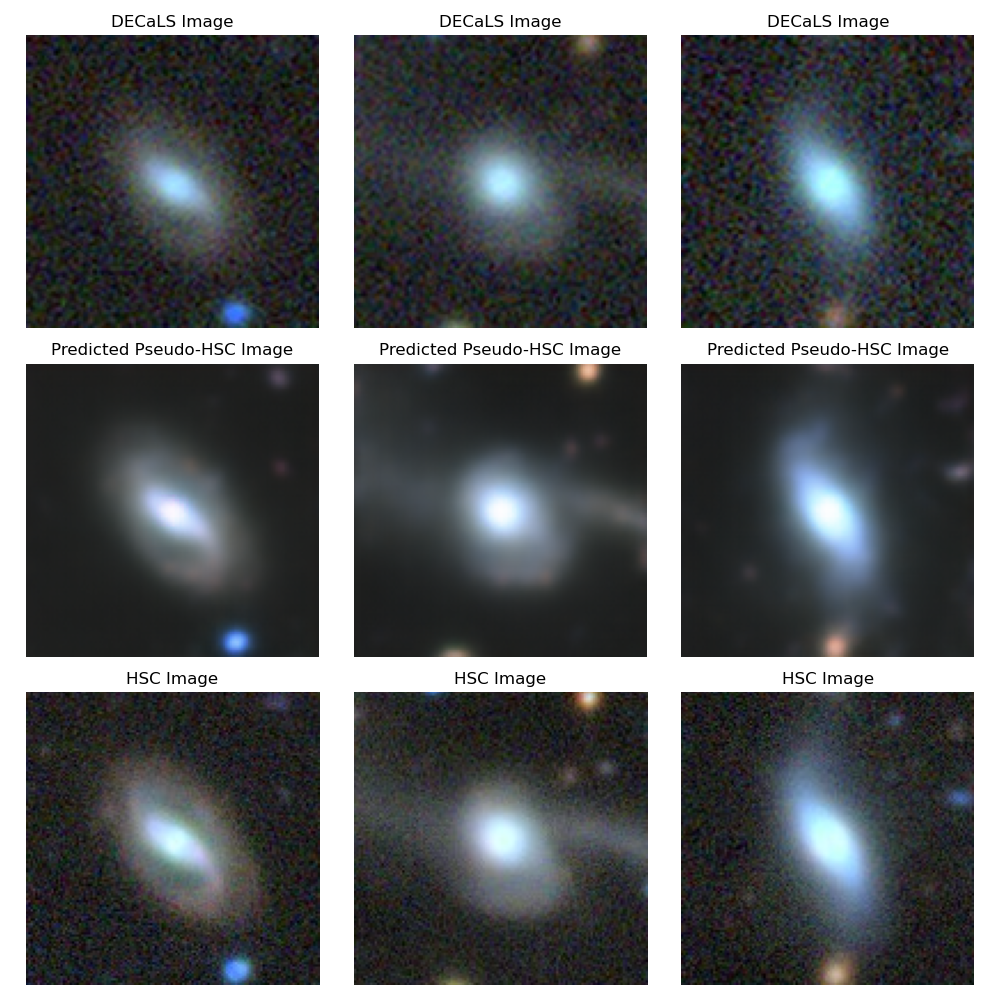}
	\caption{Example results of transforming DECaLS images to HSC images using the Pix2WGAN model. The images display three different galaxies from left to right, with the original DECaLS images shown at the top, the model-generated pseudo-HSC images in the middle, and the corresponding actual HSC observational images at the bottom.}
	\label{fig:pix2wgan_dh1}
\end{figure}

\begin{figure}
	\includegraphics[width=\textwidth]{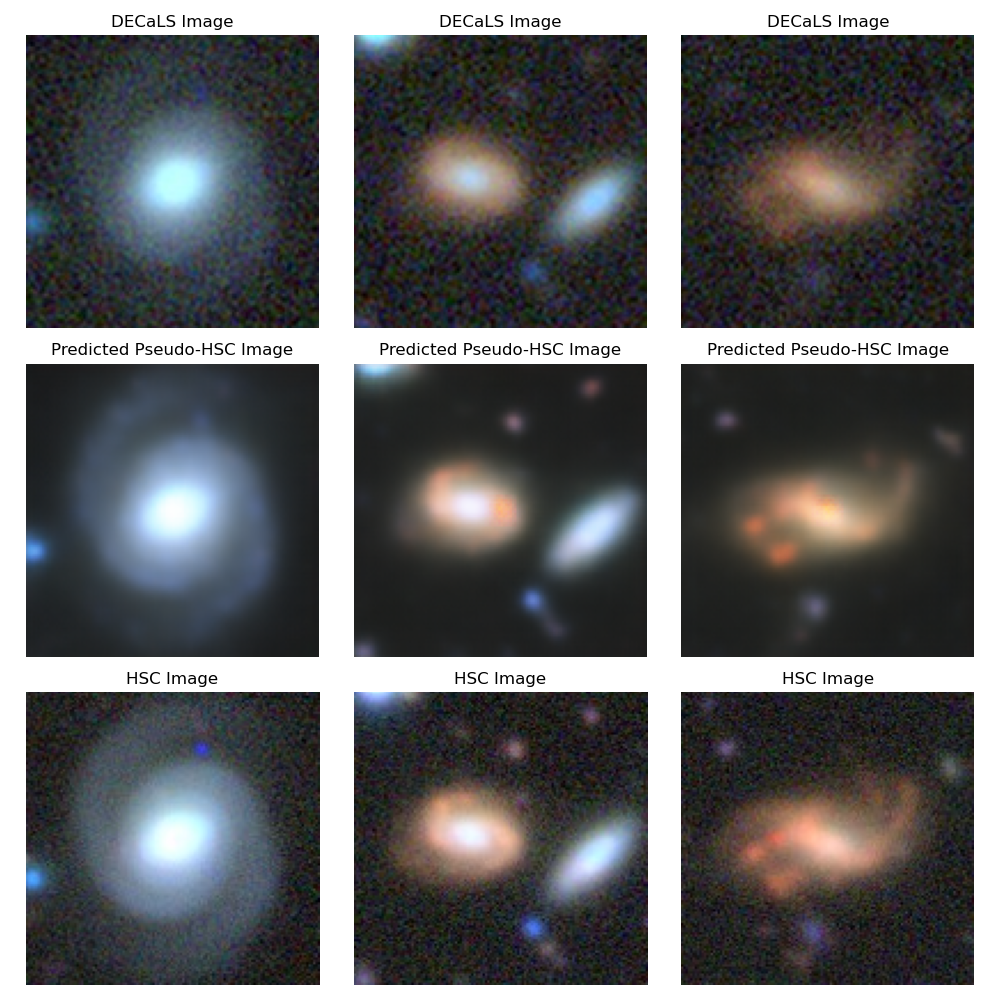}
	\caption{Similar to Figure \ref{fig:pix2wgan_dh1}. This figure presents three additional examples of transformation results from DECaLS images to HSC images using the Pix2WGAN model.}
	\label{fig:pix2wgan_dh2}
\end{figure}

For clarity, we organized the images of these selected galaxies into two figures: Figure \ref{fig:pix2wgan_dh1} and Figure \ref{fig:pix2wgan_dh2}. Each figure displays three galaxies, illustrating their images before and after transformation alongside the target images. The first row presents the original DECaLS images, the second row showcases the model-transformed pseudo-HSC images, and the third row features the corresponding actual HSC images.

Through these images, we can clearly observe the quality and detail of the pseudo-HSC images generated from various DECaLS galaxy images after the Pix2WGAN transformation. This observation validates the effectiveness of the transformation method and its ability to enhance the features and details of celestial objects. By comparing the galaxy images in Figures \ref{fig:pix2wgan_dh1} and \ref{fig:pix2wgan_dh2}, we note that the pseudo-HSC images produced by the Pix2WGAN model (second row) exhibit clearer structural features and richer details than the original DECaLS images (first row).

For example, in the first galaxy (leftmost column in both figures), the Pix2WGAN model nearly perfectly reconstructs the spiral structure with relatively smooth lines. Additionally, the disk structure of the second galaxy in Figure \ref{fig:pix2wgan_dh2} (middle column) is more pronounced. Furthermore, the tidal tail of the second galaxy in Figure \ref{fig:pix2wgan_dh1} (middle column) and the point-like structure of the third galaxy in Figure \ref{fig:pix2wgan_dh2} (rightmost column) are also clearly reproduced by the model, whereas these features are nearly invisible in the original DECaLS images. This clearly demonstrates Pix2WGAN’s exceptional performance in image transformation.

\begin{figure}
	\includegraphics[width=\textwidth]{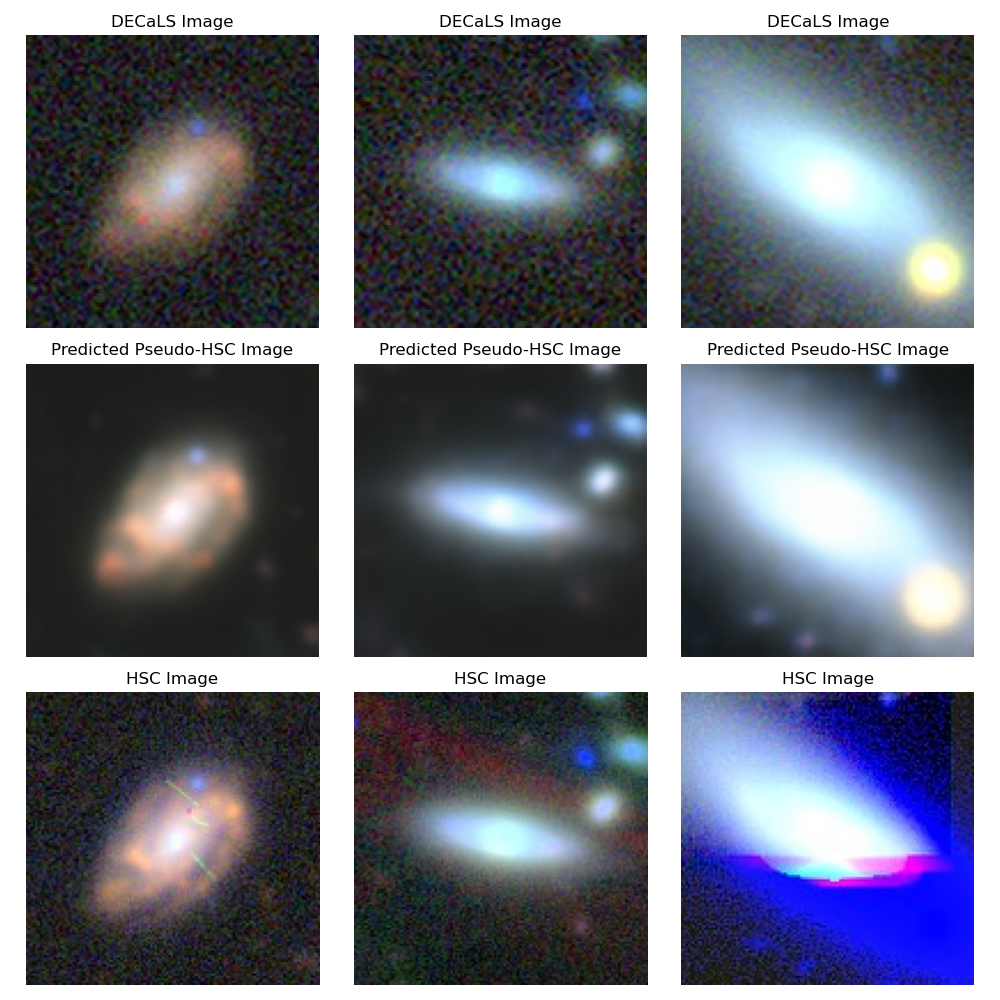}
	\caption{Similar to Figure \ref{fig:pix2wgan_dh1}. This figure shows galaxy images where the actual HSC observational images exhibit significant streak-like artifacts (left column), prominent noise (middle column), or partial overexposure (right column). In contrast, the model-transformed pseudo-HSC images successfully remove the artifacts, reduce noise interference, and correct the overexposure.}
	\label{fig:pix2wgan_noise}
\end{figure}

Furthermore, through systematic comparisons of model-reconstructed pseudo-HSC images and actual HSC observational images, we found that the Pix2WGAN model can generate images that outperform actual results in specific cases, particularly when the observational HSC images exhibit artifacts, noise, or overexposure. To illustrate this, we selected three galaxies from the test samples that display noticeable issues, as shown in Figure \ref{fig:pix2wgan_noise}. In the bottom row of the figure, the first galaxy shows streak-like artifacts, the second exhibits prominent noise, and the third is partially overexposed. After applying the Pix2WGAN model to transform the DECaLS images, the reconstructed pseudo-HSC images effectively removed artifacts, reduced noise, and corrected overexposure. 

\begin{figure}
	\includegraphics[width=\textwidth]{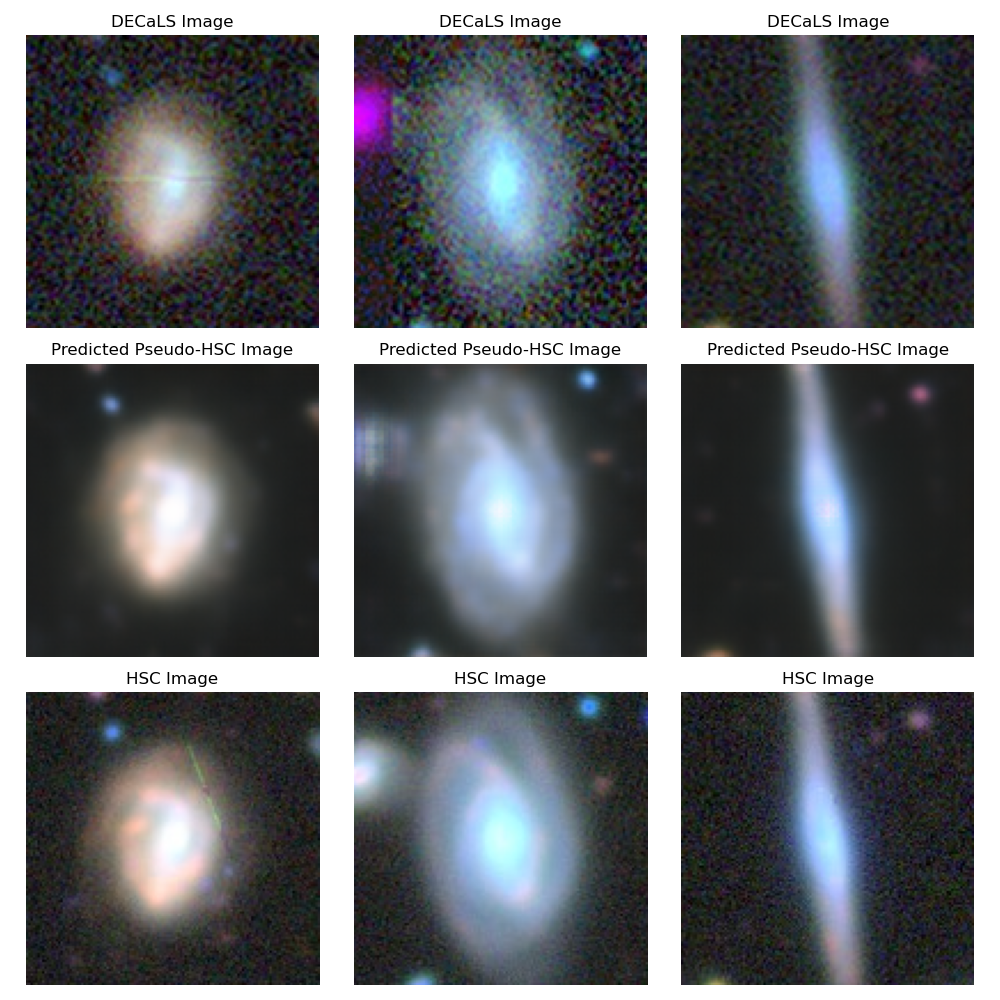}
	\caption{Similar to Figure \ref{fig:pix2wgan_dh1}. In these galaxy images, the actual DECaLS observational images exhibit significant streak-like artifacts (left column), significant noise (middle colum), , or blurriness (right column), while the model-transformed pseudo-HSC images successfully remove the artifacts, reduce noise interference, and  improve overall image clarity.}
	\label{fig:pix2wgan_noise_dec}
\end{figure}

On the other hand, we also found that when the input images contain poor-quality data (such as artifacts, significant noise, and blurriness), the pseudo-HSC images generated after transformation by our Pix2WGAN model can effectively alleviate these issues. Figure \ref{fig:pix2wgan_noise_dec} presents examples in this regard, where these galaxies exhibit clear problems in the DECaLS images: the leftmost galaxy shows noticeable streak-like artifacts, the middle galaxy has significant noise interference, and the rightmost galaxy’s image appears blurry. We observed that after the model transformation, the issues of all three galaxies showed significant improvement. Specifically, the streak-like artifacts in the leftmost DECaLS galaxy almost disappeared after transformation, the noise in the middle galaxy was greatly reduced, and the blurriness in the rightmost galaxy’s image was also alleviated to some extent. 

In fact, a careful analysis of Figures \ref{fig:pix2wgan_dh1} and \ref{fig:pix2wgan_dh2} also shows that the Pix2WGAN model effectively reduces background noise during the generation of pseudo-HSC images. These results further highlight the potential of the Pix2WGAN model in addressing issues related to image artifacts, noise, overexposure, and blurriness. Particularly when the quality of observational images is limited or flawed, the model demonstrates the ability to provide effective image enhancement and repair capabilities.

However, it is important to note that the model-predicted pseudo-HSC images (second row) shown in Figures \ref{fig:pix2wgan_dh1} and \ref{fig:pix2wgan_dh2} exhibit similarities in structure and detail to the actual HSC observational images (third row); however, they still fall short in overall clarity and detail representation. This suggests that while deep learning models can significantly enhance the quality of the original images, they cannot completely replace the need for hardware upgrades and improvements. Therefore, the most effective strategy for improving image quality is to combine advanced deep learning techniques with high-performance hardware.

\subsection{Transforming Images from SDSS to HSC} \label{subsection:SDSS_HSC}

%In this subsection, we examine the image transformation from SDSS to HSC. There is a significant difference in the quality of observational images produced by these two survey projects. For example, the pixel resolution of SDSS images is 0.396 arcseconds per pixel, while HSC images have a resolution of 0.168 arcseconds per pixel, resulting in a resolution advantage of 2.357 times. This difference is greater than the approximate 1.511 times difference between DECaLS and HSC, making the direct application of the Pix2WGAN model for transforming SDSS images to HSC images less than ideal.

In this subsection, we explore the image transformation from SDSS to HSC. Compared to the transformation from DECaLS to HSC, there is a significant difference in the quality of the observed images produced by the SDSS and HSC survey projects. For example, the pixel resolution of SDSS images is 0.396 arcseconds per pixel, while HSC images have a resolution of 0.168 arcseconds per pixel, indicating that HSC images possess approximately 2.357 times the resolution advantage. This disparity is notably greater than the roughly 1.511 times difference between DECaLS and HSC, which makes the direct application of the Pix2WGAN model for transforming SDSS images to HSC images less than optimal.

To achieve better results in image transformation, we further enhanced the architecture of the hybrid model Pix2WGAN, allowing it to utilize intermediate images as a bridge. These bridge images can provide additional information, significantly improving the stability and accuracy of the generated results. These intermediate images can be either DECaLS images or degraded HSC images (e.g., reducing the resolution of HSC images by approximately half). In this study, we have chosen DECaLS images as the bridging images, enabling us to simultaneously incorporate images from SDSS, DECaLS, and HSC for cascade training. We refer to this modified model as the Cascade Pix2WGAN model. 

\begin{figure}
	\includegraphics[width=\textwidth]{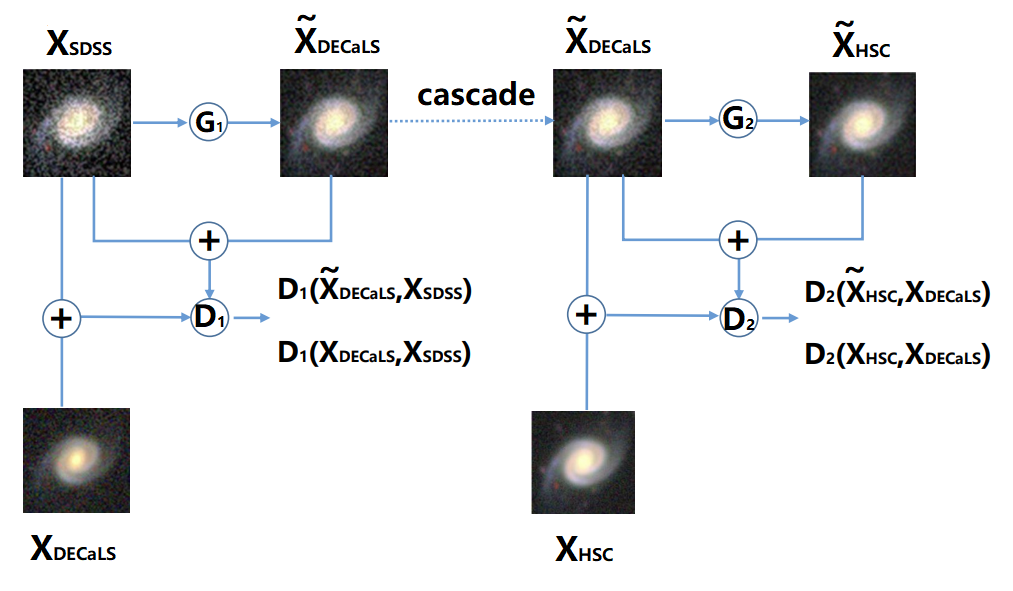}
	\caption{Architecture of the Cascade Pix2WGAN Model. $G_1$ and $G_2$ represent the generators, while $D_1$ and $D_2$ represent the discriminators. $\rm{X}_{SDSS}$, $\rm{X}_{DECaLS}$, and $\rm{X}_{HSC}$ denote the real observational images obtained from SDSS, DECaLS, and HSC, respectively. Meanwhile, $\widetilde{\rm{X}}_{\rm{DECaLS}}$ and $\widetilde{\rm{X}}_{\rm{HSC}}$ represent the pseudo-DECaLS and pseudo-HSC images generated by the model, respectively.}
	\label{fig:cascade_archi}
\end{figure}

It should be noted that our model does not have strict criteria for limiting or stopping the use of bridge images; whether to use bridge images primarily depends on the quality requirements of the generated images. When the demand for image quality is high, we recommend using bridge images whenever possible, as they can provide additional pixel information that helps the model generate higher-quality target images. Our test results show that even when not using bridge images, the model can directly generate pseudo-HSC images from SDSS images; however, the quality and effectiveness of the generated images are generally lower without bridge images. The Cascade architecture of the model is illustrated in Figure \ref{fig:cascade_archi}, and the detailed training process is described as follows:

First, the SDSS images are input into the Cascade Pix2WGAN model, which generates pseudo-DECaLS images using the first generator, $G_1$. Next, the pseudo-DECaLS images generated by $G_1$ are paired with the original SDSS images to form one set, while another set consists of real DECaLS images along with the original SDSS images. Both sets of images are then input into the first discriminator, $D_1$, whose task is to measure the difference between the real and fake images in these sets.

Subsequently, the pseudo-DECaLS images generated by $G_1$ are fed into the second generator, $G_2$, to produce pseudo-HSC images. The pseudo-HSC images are paired with their corresponding pseudo-DECaLS images to form one set, while another set consists of the corresponding real HSC images and pseudo-DECaLS images. These two sets of images are then input into the second discriminator, $D_2$, which learns to measure the difference between the real and fake images in this context.

All the generators and discriminators in the Cascade Pix2WGAN model share the same structures and parameter settings as described in Section \ref{subsection:DECaLS_HSC}. In each iteration of every training epoch, the generators $G_1$ and $G_2$, along with the discriminators $D_1$ and $D_2$, are optimized sequentially based on their interdependencies, with the goal of collectively improving the performance of the model. After multiple iterations, the final trained model can directly transform the original SDSS images into pseudo-HSC images using generators $G_1$ and $G_2$, without the need for DECaLS images as an intermediate bridge. The entire training process of the Cascade Pix2WGAN model took approximately 13.2 hours.

\begin{figure}
	\includegraphics[width=\textwidth]{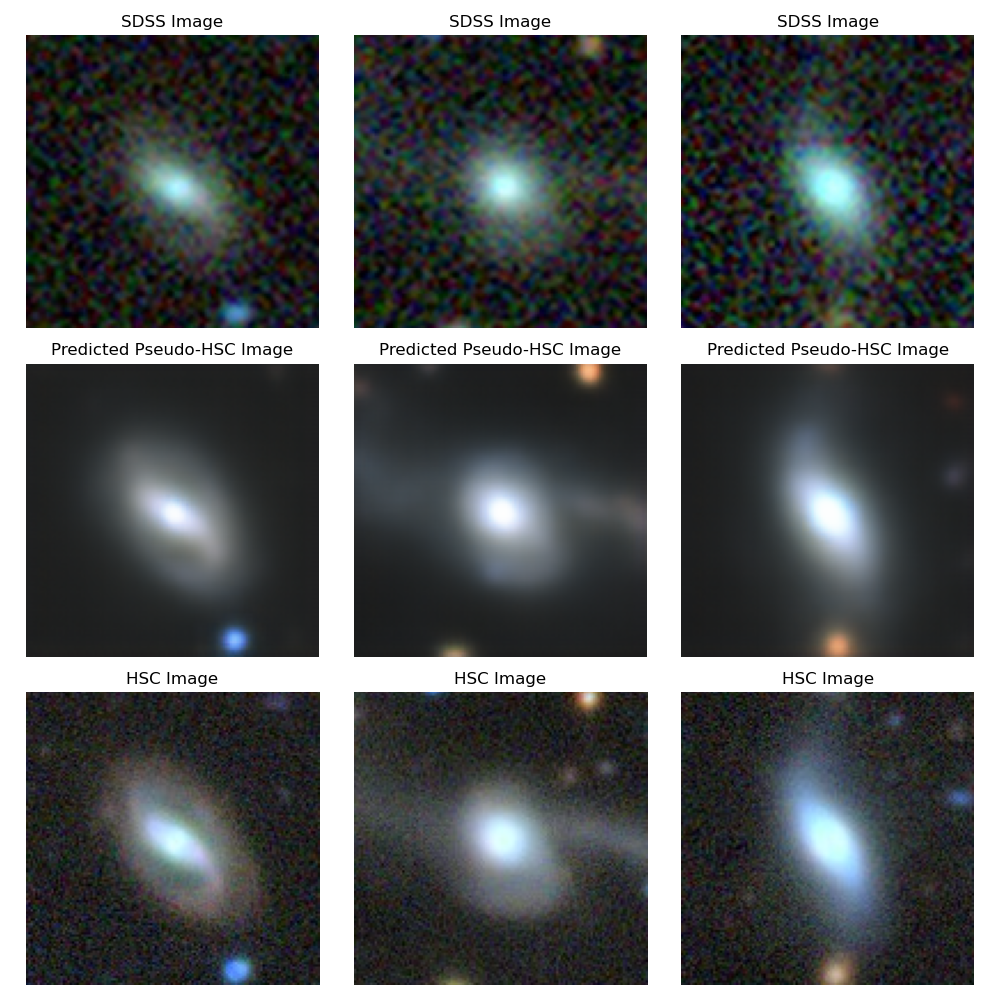}
	\caption{Example results of transforming SDSS images to HSC images using the Cascade Pix2WGAN model. The images display the same three galaxies as in Figure \ref{fig:pix2wgan_dh1}, with the original SDSS images shown at the top, the model-generated pseudo-HSC images in the middle, and the corresponding actual HSC observational images at the bottom.}
	\label{fig:pix2wgan_sdh1}
\end{figure}

\begin{figure}
	\includegraphics[width=\textwidth]{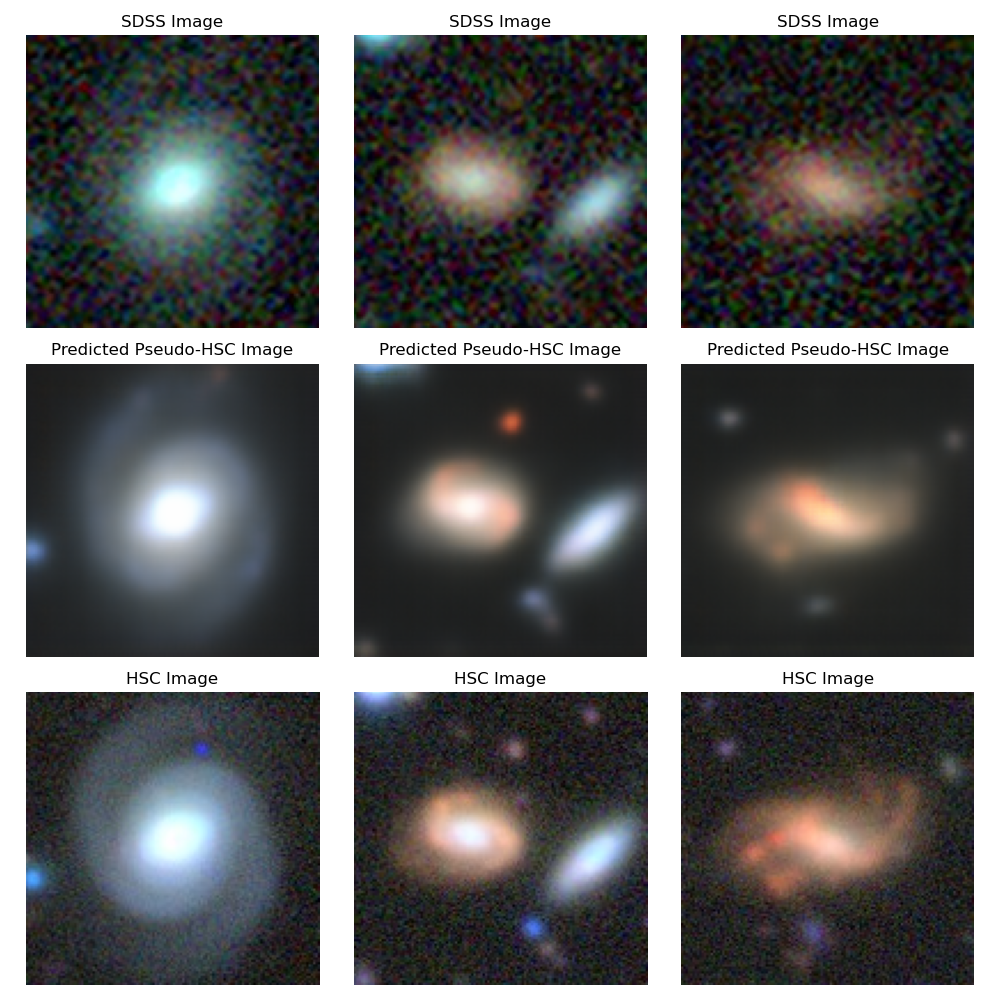}
	\caption{Similar to Figure \ref{fig:pix2wgan_sdh1}. The images display the same three galaxies as in Figure \ref{fig:pix2wgan_dh2}, with the original SDSS images shown at the top, the model-generated pseudo-HSC images in the middle, and the corresponding actual HSC observational images at the bottom.}
	\label{fig:pix2wgan_sdh2}
\end{figure}

To facilitate comparison with the previously mentioned results, we selected six galaxies from the test set that are the same as those shown in Figures \ref{fig:pix2wgan_dh1} and \ref{fig:pix2wgan_dh2}. The transformation effects of the cascade model are displayed in Figures \ref{fig:pix2wgan_sdh1} and \ref{fig:pix2wgan_sdh2}. Each figure showcases the transformation visual effects of the three galaxies: the top row displays the original SDSS images, the middle row presents the pseudo-HSC images generated by the Cascade Pix2WGAN model, and the bottom row shows the corresponding real HSC images.

From Figures \ref{fig:pix2wgan_sdh1} and \ref{fig:pix2wgan_sdh2}, it is evident that our Cascade Pix2WGAN model has achieved remarkable results. Many structural features that are not visible to the naked eye in SDSS—such as spiral arms, disks, tidal tails, bridges, and fine structures near the centers of galaxies—are clearly presented. The pseudo-HSC images generated by the model closely resemble the actual observed HSC images in terms of galaxy structure, morphology, and color. Compared to the original SDSS images, the transformed images exhibit significant improvements in resolution and detail recognition capability. We believe that these transformed SDSS images will serve as reliable data sources for future research in galaxy morphology classification and structural feature recognition. Additionally, as noted in the previous section, the generated pseudo-HSC images effectively reduce background noise, thereby improving the signal-to-noise ratio of the galaxy images.

Furthermore, we compared the middle row panels of Figures \ref{fig:pix2wgan_sdh1} and \ref{fig:pix2wgan_sdh2} with those in Figures \ref{fig:pix2wgan_dh1} and \ref{fig:pix2wgan_dh2}. The results indicate that the pseudo-HSC images generated directly from SDSS images using the Cascade Pix2WGAN model perform slightly worse than those generated from DECaLS images using the Pix2WGAN model. This observation suggests that while the model’s performance is important, the quality of the initial input images is even more critical. We will explore this point in more detail in our subsequent quantitative comparisons.

\subsection{Quantitative Metrics Assessment} \label{secsection:metrics}

In sections \ref{subsection:DECaLS_HSC} and \ref{subsection:SDSS_HSC}, we conducted a visual inspection of the generated pseudo-HSC images. The results indicated that the Pix2WGAN model significantly enhances the quality of SDSS and DECaLS images, and the generated pseudo-HSC images exhibit a high degree of similarity to the real HSC images. To further validate the model’s performance, this subsection introduces several quantitative evaluation metrics to assess the differences between the generated pseudo-HSC images and the real HSC images. These metrics include Root Mean Squared Error (RMSE), Peak Signal-to-Noise Ratio (PSNR), and Structural Similarity Index (SSIM; \citealt{sheikh2004wang}), as well as two perceptual metrics: Learned Perceptual Image Patch Similarity (LPIPS; \citealt{zhang2018unreasonable}) and Fréchet Inception Distance (FID; \citealt{heusel2017gans}).

RMSE is a commonly used image similarity metric that measures the degree of difference between two images. This metric is calculated by averaging the squared differences between corresponding pixels of the two images  (known as Mean Squared Error, MSE) and then taking the square root to obtain a similarity score. A smaller RMSE value indicates greater similarity between the two images, with a value of 0 representing that they are identical. For two multi-channel images of size $M \times N \times C$, the calculation formula is:
\begin{equation}
    \text{RMSE} = \sqrt{\frac{1}{M \times N \times C} \sum_{i=1}^{M} \sum_{j=1}^{N} \sum_{k=1}^{C} (I_{ijk} - K_{ijk})^2} ,
	\label{eq:mse}
\end{equation}
where $I$ is the predicted image generated by the model and $K$ is the real image.

PSNR is also an important metric widely used to evaluate the performance of various image processing algorithms. Its value is derived from the MSE, establishing an interdependent relationship between the two metrics. PSNR is typically expressed in decibels (dB), where a higher value indicates greater similarity between the generated image and the real image, as well as better image quality. The calculation formula for PSNR is as follows:

\begin{equation}
  \text{PSNR} = 10 \log_{10} \left( \frac{MAX^2}{\text{MSE}} \right) ,
  \label{eq:psnr}
\end{equation}

where $MAX$ represents the maximum possible pixel value in the image; in this paper, the maximum pixel value is 255.

Unlike RMSE and PSNR, which evaluate the quality of generated images based on pixel differences, SSIM measures the similarity between the real image and the generated image by comparing their brightness, contrast, and structural information. In contrast to RMSE and PSNR, SSIM places greater emphasis on structural similarity rather than on pixel-by-pixel brightness differences, making it a more accurate reflection of human visual perception of image similarity. The SSIM value ranges from [-1, 1], with higher values indicating greater similarity between the two images. Its calculation involves comparing multiple features, including mean, variance, covariance, and two constants used for stability, as shown in the following formula:

\begin{equation}
  \text{SSIM}(x, y) = \frac{(2\mu_x\mu_y + C_1)(2\sigma_{xy} + C_2)}{(\mu_x^2 + \mu_y^2 + C_1)(\sigma_x^2 + \sigma_y^2 + C_2)} .
  \label{eq:ssim}
\end{equation}
In the formula, $x$ and $y$ represent the generated image and the real image, respectively, while $\mu_x$, $\mu_y$, $\sigma_x^2$, $\sigma_y^2$, and $\sigma_{xy}$ represent the mean, variance, and covariance of the generated image $x$ and the real image $y$. $C_1 = (K_1 L)^2$ and $C_2 = (K_2 L)^2$, where $L$ is the dynamic range of pixel values, and $K1$ and $K2$ are small constants used to ensure stability in calculations. In this study, $L = 255$, $K_1 = 0.01$, and $K_2 = 0.03$.

In addition, to better assess the differences between generated images and real images from the perspective of the human visual system, we introduce two commonly used metrics in deep learning: LPIPS and FID. These metrics evaluate image differences based on features extracted by deep learning models, enabling a more effective capture of perceptual similarity. Smaller values of LPIPS and FID indicate higher perceptual similarity between the two images. For detailed definitions and calculation methods of LPIPS and FID, please refer to the studies by \citet{zhang2018unreasonable} and \citet{szegedy2015going}. Compared to traditional image quality assessment methods, these metrics provide a more comprehensive analysis and complement the evaluation of generated image quality.

Using these evaluation metrics, we quantitatively assessed the similarity between the model-generated pseudo-HSC images and the real HSC observational images, as well as the similarity between the pre-transformed images and the real HSC images. The results for the test set are presented in Table \ref{tab:metrics}. In the table, the superscript “1” after the term “pseudo-HSC” indicates that these images were generated from DECaLS images using the Pix2WGAN model, while the superscript “2” denotes that the images were generated from SDSS images using the Cascaded Pix2WGAN model.

\begin{table}[h]
\centering
\caption{Comparison of similarity assessment metrics for different paired samples. The superscript “1” indicates images generated from DECaLS using the Pix2WGAN model, while “2” indicates images generated from SDSS using the Cascaded Pix2WGAN model. \label{tab:metrics}}
\begin{tabular}{lccccc}
\hline
Image Pairs       & RMSE  & PSNR  & SSIM & LPIPS & FID    \\ \hline
SDSS -- HSC       & 34.81 & 17.59 & 0.43 & 0.19  & 179.59 \\
Pseudo-HSC$^{2}$ -- HSC & 21.69 & 23.19 & 0.71 & 0.11  & 105.16 \\
DECaLS -- HSC     & 27.39 & 19.95 & 0.56 & 0.06  & 78.91  \\
Pseudo-HSC$^{1}$ -- HSC & 20.04 & 24.21 & 0.72 & 0.09  & 98.46  \\ \hline
\end{tabular}
\end{table}

From Table 1, it is evident that the similarity between the pseudo-HSC images generated by the Cascaded Pix2WGAN model and the real HSC images is significantly higher across all metrics compared to the similarity between the original SDSS images and the real HSC images. Specifically, the SSIM measure increased significantly from 0.43 to 0.71, PSNR improved from 17.59 dB to 23.19 dB, and RMSE decreased from 34.81 to 21.69. At the same time, the LPIPS and FID metrics also showed substantial improvement, with LPIPS decreasing from 0.19 to 0.11 and FID dropping from 179.59 to 105.16.

Similarly, the similarity between the pseudo-HSC images generated by the Pix2WGAN model and the real HSC images is significantly higher on most metrics compared to the similarity between the original DECaLS images and the real HSC images. Specifically, the SSIM metric improved from 0.56 to 0.72, PSNR increased from 19.95 dB to 24.21 dB, and RMSE decreased from 27.39 to 20.04. This indicates that the Pix2WGAN model excels in reconstructing the details and optical quality of the images.

However, it is worth noting that the LPIPS and FID metrics showed limited change, with LPIPS rising slightly from 0.06 to 0.09 and FID increasing from 78.91 to 98.46. This may be because both LPIPS and FID rely on pre-trained deep learning models (such as VGG and Inception) to capture image features and assess similarity between images, and the performance of these models largely depends on the datasets on which they were trained. Generally, the pre-trained models for LPIPS and FID are trained on the ImageNet dataset, which encompasses a wide variety of objects and scenes but does not include astronomical images. Therefore, when evaluating galaxy images, these models may not effectively capture the unique structures, features, and visual information specific to galaxy images, leading to small differences in LPIPS and FID that do not accurately reflect the similarity or differences between galaxy images.

Nonetheless, the overall evaluation still indicates that the Pix2WGAN model effectively enhances image quality, particularly in terms of preserving structural integrity and visual information.

Furthermore, the results in Table \ref{tab:metrics} indicate that the pseudo-HSC images generated from DECaLS images using the Pix2WGAN model outperform those generated from SDSS images with the Cascaded Pix2WGAN model across all evaluation metrics. This suggests that while optimizing the performance of the model itself is crucial for the generated results, the quality of the original input images should not be overlooked. High-quality input images provide the model with richer details and more accurate information, significantly enhancing the transformation results. This is also visually evident in the comparisons between Figures \ref{fig:pix2wgan_dh1}, \ref{fig:pix2wgan_dh2}, and Figures \ref{fig:pix2wgan_sdh1}, \ref{fig:pix2wgan_sdh2}: the pseudo-HSC images generated from DECaLS images show superior performance in detail restoration, signal-to-noise ratio, and clarity compared to the results generated from SDSS images.

\section{Summary} \label{section:Summary}

%In recent years, deep learning technology has made significant progress in image enhancement tasks and has been widely applied in astrophysics. However, these technologies primarily focus on denoising astronomical optical or radio images (Lin et al., 2021; Vojtekova et al., 2021; Liu et al., 2023) and on super-resolution reconstruction of astronomical images (Li et al., 2022; Peng et al., 2023; Shibuya et al., 2024; Miao et al., 2024). In contrast, the use of deep learning techniques for image conversion between different telescopes remains a relatively underexplored area.

%This paper investigates image conversion between different survey projects, specifically examining how to utilize images from the Sloan Digital Sky Survey (SDSS) and the Dark Energy Camera Legacy Survey (DECaLS) to generate galaxy images of quality comparable to those from the Subaru/Hyper Suprime-Cam Survey (HSC). To achieve this, we propose a hybrid model—Pix2WGAN—that combines the strengths of pix2pix and the Wasserstein GAN with gradient penalty (WGAN-GP) to effectively convert low-quality observational images into high-quality ones. We first successfully applied the model to convert DECaLS images to HSC images. Subsequently, for the significantly different SDSS and HSC images, we introduced a novel cascaded Pix2WGAN model architecture. This architecture employs a multi-stage training mechanism to effectively address substantial differences in resolution, dynamic range, and signal-to-noise ratio among these images, thereby enhancing performance in converting datasets with marked disparities.

In recent years, deep learning technology has made significant advancements in image enhancement tasks and has been widely applied in the field of astrophysics. These technologies primarily focus on the denoising of astronomical images \citep{lin2021galaxy,vojtekova2021learning} and super-resolution reconstruction \citep{li2022self,shibuya2024galaxy,miao2024astrosr}. By optimizing noise suppression and enhancing resolution, these methods have substantially improved the quality of astronomical images, allowing for clearer presentations of finer details and thereby supporting more precise astronomical analysis. However, the application of deep learning techniques to image transformation between different telescopes remains a relatively underexplored area.

Compared to denoising and super-resolution tasks, image transformation between telescopes involves not only issues of image resolution and noise processing, but also significant variations in the imaging characteristics of different telescopes, observing conditions, and data distributions. This transformation method can elevate low-resolution, low signal-to-noise ratio images to higher quality standards, thereby enhancing the scientific value of the data and more efficiently supporting astronomical analysis and research. Although research in this area is still limited, the rapid development of deep learning technology is gradually revealing the potential for cross-telescope image transformation, which is expected to become an important tool in astrophysics in the future.

In this study, we explored galaxy image transformation between different astronomical survey projects, specifically utilizing images from SDSS and DECaLS to generate images that are comparable to the high-quality galaxy images obtained from HSC observations. This work serves as a proof of concept for the validity of using deep learning techniques for this type of application. To achieve this, we proposed a hybrid model—Pix2WGAN—that combines the strengths of both pix2pix and WGAN-GP, enabling effective transformation of low-quality observational images into high-quality ones. We successfully applied this model to transform DECaLS images into pseudo-HSC images, achieving promising results.

Building on this, we further introduced a novel Cascaded Pix2WGAN model architecture specifically designed to address the significant differences between SDSS and HSC images. This architecture employs a multi-stage training mechanism that effectively manages the differences in resolution, dynamic range, and signal-to-noise ratio between the two types of images, significantly enhancing the model’s performance when transforming datasets with marked disparities. Through this design, the model systematically bridges the gap between low-quality SDSS images and high-quality HSC images, resulting in the generation of superior pseudo-HSC images.

%This paper examines image transformation between different astronomical survey projects, particularly how to utilize images from SDSS and DECaLS to generate galaxy images comparable to those obtained from HSC observations. To this end, we propose a hybrid model—Pix2WGAN—that combines the strengths of pix2pix and WGAN-GP to effectively transform low-quality observational images into high-quality images. We first successfully applied this model to transform DECaLS images into pseudo-HSC images. Building on this, we introduced a novel cascaded Pix2WGAN model architecture to address the significant differences between SDSS and HSC images. This architecture adopts a multi-stage training mechanism that effectively resolves the differences in resolution, dynamic range, and signal-to-noise ratio between the two types of images, thereby enhancing the model’s performance in transforming datasets with marked disparities.

To validate the quality of the images generated by the Pix2WGAN and Cascaded Pix2WGAN models, we employed both qualitative and quantitative assessment methods to compare the model-generated pseudo-HSC images with their corresponding real HSC images. This evaluation not only confirmed the reliability of the generated images but also provided a comprehensive analysis of the model's performance in detail recovery and overall quality enhancement. The qualitative assessment involved visual inspection, which demonstrated that the overall quality of the generated images significantly exceeded that of the initial input images, particularly regarding galaxy structure, internal details, and noise levels. The generated pseudo-HSC images closely approximated the real HSC images in terms of detail richness and clarity, showcasing the model’s exceptional capability in structural representation.

In the quantitative evaluation, we measured the similarity between the model-generated images and the actual observed images by calculating various commonly used metrics. Specifically, we selected RMSE, PSNR, SSIM, and two perceptual quality metrics: LPIPS and FID, to quantitatively assess the quality of the generated images. The results indicated that the pseudo-HSC images generated by the model performed excellently across most metrics, with the differences between them and the actual HSC observed images being significantly smaller than those between the initial input images and the real HSC observed images. This demonstrates that our constructed image transformation model can significantly improve the quality of SDSS and DECaLS images, producing pseudo-HSC images that closely resemble the actual HSC images.

Upon further inspection and analysis, we found that our model effectively addresses issues associated with low-quality input data, such as artifacts, significant noise, and blurriness. Furthermore, the model demonstrates the capability to generate images that surpass the quality of actual observed images under certain conditions. For instance, when HSC images contain artifacts, noise, or overexposure, the model can produce clearer and smoother images, thereby mitigating the impact of these observational errors. This finding indicates that the Pix2WGAN model possesses a degree of fault tolerance while enhancing image quality, allowing it to correct imaging defects caused by instrumental or environmental factors present in actual observations.

However, it is important to note that, while the Pix2WGAN model can significantly improve the quality of input images, the generated pseudo-HSC images still exhibit certain differences from real HSC images in terms of overall clarity and detail. This indicates that the generative model has not yet been able to capture small features with the same precision as hardware observations, resulting in issues such as image blurriness and discrepancies in feature sizes. Therefore, caution should be exercised when using single images generated by the model for detailed astrophysical measurements. However, by performing statistical analyses on a large number of generated galaxy images, we can effectively reduce the random biases introduced by individual images, thereby obtaining more reliable statistical results. {In our future work, we aim to address these limitations by refining our model architecture and incorporating additional training data, which will enhance the accuracy and precision of the generated images. Additionally, we plan to integrate more advanced astrophysical indicators into our analyses, as suggested by \citet{wang2023neural}, to provide a more comprehensive evaluation of our method’s applicability in astrophysical research.}
%Further improvements in image transformation could be achieved by increasing the number of training samples and adopting more complex model architectures.

In addition, although deep learning models demonstrate considerable potential in image transformation and enhancement, they cannot fully replace advanced hardware upgrades and improvements. Future efforts in image processing and enhancement should integrate advanced deep learning technologies with high-performance observational equipment to collectively elevate the quality of astronomical images.

Our research has enabled the construction of a pseudo-HSC dataset within the observational regions of the SDSS and DECaLS, covering areas of the sky not yet observed by HSC, thus increasing the utilization efficiency of existing low-quality astronomical data. Additionally, the proposed model is not only applicable to image transformation between the SDSS, DECaLS, and HSC datasets, but can also be extended to other survey projects, such as the Dark Energy Survey (DES), the Large Synoptic Survey Telescope (LSST), the Hubble Space Telescope (HST), the Chinese Space Station Telescope(CSST) and the James Webb Space Telescope (JWST). In future applications, this method is anticipated to offer valuable technical support for astrophysical data analysis, facilitate image integration across different survey projects, and enable high-precision astrophysical measurements.

%% IMPORTANT! The old "\acknowledgment" command has be depreciated. It was
%% not robust enough to handle our new dual anonymous review requirements and
%% thus been replaced with the acknowledgment environment. If you try to 
%% compile with \acknowledgment you will get an error print to the screen
%% and in the compiled pdf.
%% 
%% Also note that the akcnowlodgment environment does not support long amounts of text. If you have a lot of people and institutions to acknowledge, do not use this command. Instead, create a new \section{Acknowledgments}.
%\begin{acknowledgments}
\section{Acknowledgments}

Z.J.L. acknowledges the support from the Shanghai Science and Technology Foundation Fund (Grant No. 20070502400) and the scientific research grants from the China Manned Space Project. S.H.Z. acknowledges support from the National Natural Science Foundation of China (Grant No. 12173026), the National Key Research and Development Program of China (Grant No. 2022YFC2807303), the Shanghai Science and Technology Fund (Grant No. 23010503900), the Program for Professor of Special Appointment (Eastern Scholar) at Shanghai Institutions of Higher Learning, and the Shuguang Program (23SG39) of the Shanghai Education Development Foundation and Shanghai Municipal Education Commission. L.P.F. acknowledges the support from the National Natural Science Foundation of China (NSFC 11933002). H.B.X. acknowledges the support from the National Natural Science Foundation of China (NSFC 12203034) and the Shanghai Science and Technology Fund (22YF1431500). W.D. acknowledges the support from NSFC Grant No. 11890691. This work is also supported by the National Natural Science Foundation of China under Grant No. 12141302.

The Hyper Suprime-Cam (HSC) collaboration includes the astronomical communities of Japan and Taiwan, as well as Princeton University. The HSC instrumentation and software were developed by the National Astronomical Observatory of Japan (NAOJ), the Kavli Institute for the Physics and Mathematics of the Universe (Kavli IPMU), the University of Tokyo, the High Energy Accelerator Research Organization (KEK), the Academia Sinica Institute for Astronomy and Astrophysics in Taiwan (ASIAA), and Princeton University. Funding was contributed by various organizations including the Japanese Cabinet Office, the Ministry of Education, Culture, Sports, Science and Technology (MEXT), and the Japan Society for the Promotion of Science (JSPS), and others.

The DESI Legacy Imaging Surveys consist of three individual projects: the Dark Energy Camera Legacy Survey (DECaLS), the Beijing-Arizona Sky Survey (BASS), and the Mayall z-band Legacy Survey (MzLS). These surveys utilized facilities such as the Blanco, Bok, and Mayall telescopes, supported by the National Science Foundation (NSF) and operated by different observatories including NSF’s NOIRLab.

Funding for the Sloan Digital Sky Survey has been provided by the Alfred P. Sloan Foundation, the Heising-Simons Foundation, the National Science Foundation, and participating institutions. SDSS telescopes are located at Apache Point Observatory and Las Campanas Observatory.

We thank the respective teams and funding agencies for making these data publicly available. For detailed acknowledgments and funding information, please refer to the original publications and data release notes.

%\end{acknowledgments}

%% To help institutions obtain information on the effectiveness of their 
%% telescopes the AAS Journals has created a group of keywords for telescope 
%% facilities.
%
%% Following the acknowledgments section, use the following syntax and the
%% \facility{} or \facilities{} macros to list the keywords of facilities used 
%% in the research for the paper.  Each keyword is check against the master 
%% list during copy editing.  Individual instruments can be provided in 
%% parentheses, after the keyword, but they are not verified.

%\vspace{5mm}
%\facilities{HST(STIS), Swift(XRT and UVOT), AAVSO, CTIO:1.3m,
%CTIO:1.5m,CXO}

%% Similar to \facility{}, there is the optional \software command to allow 
%% authors a place to specify which programs were used during the creation of 
%% the manuscript. Authors should list each code and include either a
%% citation or url to the code inside ()s when available.

%\software{astropy \citep{2013A&A...558A..33A,2018AJ....156..123A},  
%          Cloudy \citep{2013RMxAA..49..137F}, 
%          Source Extractor \citep{1996A&AS..117..393B}
%          }

%% For this sample we use BibTeX plus aasjournals.bst to generate the
%% the bibliography. The sample631.bib file was populated from ADS. To
%% get the citations to show in the compiled file do the following:
%%
%% pdflatex sample631.tex
%% bibtext sample631
%% pdflatex sample631.tex
%% pdflatex sample631.tex

\bibliography{ref}{}

\begin{thebibliography}{}
\expandafter\ifx\csname natexlab\endcsname\relax\def\natexlab#1{#1}\fi
\providecommand{\url}[1]{\href{#1}{#1}}
\providecommand{\dodoi}[1]{doi:~\href{http://doi.org/#1}{\nolinkurl{#1}}}
\providecommand{\doeprint}[1]{\href{http://ascl.net/#1}{\nolinkurl{http://ascl.net/#1}}}
\providecommand{\doarXiv}[1]{\href{https://arxiv.org/abs/#1}{\nolinkurl{https://arxiv.org/abs/#1}}}

\bibitem[{Abadi {et~al.}(2016)Abadi, Barham, Chen, Chen, Davis, Dean, Devin, Ghemawat, Irving, Isard, {et~al.}}]{abadi2016tensorflow}
Abadi, M., Barham, P., Chen, J., {et~al.} 2016, in 12th USENIX symposium on operating systems design and implementation (OSDI 16), 265--283

\bibitem[{Abazajian {et~al.}(2009)Abazajian, Adelman-McCarthy, Ag{\"u}eros, Allam, Prieto, An, Anderson, Anderson, Annis, Bahcall, {et~al.}}]{abazajian2009seventh}
Abazajian, K.~N., Adelman-McCarthy, J.~K., Ag{\"u}eros, M.~A., {et~al.} 2009, The Astrophysical Journal Supplement Series, 182, 543

\bibitem[{Adler \& Lunz(2018)}]{adler2018banach}
Adler, J., \& Lunz, S. 2018, Advances in neural information processing systems, 31

\bibitem[{Ahumada {et~al.}(2020)Ahumada, Prieto, Almeida, Anders, Anderson, Andrews, Anguiano, Arcodia, Armengaud, Aubert, {et~al.}}]{ahumada202016th}
Ahumada, R., Prieto, C.~A., Almeida, A., {et~al.} 2020, The Astrophysical Journal Supplement Series, 249, 3

\bibitem[{Arcelin {et~al.}(2021)Arcelin, Doux, Aubourg, Roucelle, \& Collaboration)}]{arcelin2021deblending}
Arcelin, B., Doux, C., Aubourg, E., Roucelle, C., \& Collaboration), L. D. E.~S. 2021, Monthly Notices of the Royal Astronomical Society, 500, 531

\bibitem[{Bai {et~al.}(2008)Bai, Bajaj, Beletic, Farris, Joshi, Lauxtermann, Petersen, \& Williams}]{bai2008teledyne}
Bai, Y., Bajaj, J., Beletic, J.~W., {et~al.} 2008, in High energy, optical, and infrared detectors for astronomy III, Vol. 7021, SPIE, 29--44

\bibitem[{Blanton {et~al.}(2011)Blanton, Kazin, Muna, Weaver, \& Price-Whelan}]{blanton2011improved}
Blanton, M.~R., Kazin, E., Muna, D., Weaver, B.~A., \& Price-Whelan, A. 2011, The Astronomical Journal, 142, 31

\bibitem[{Boucaud {et~al.}(2020)Boucaud, Huertas-Company, Heneka, Ishida, Sedaghat, de~Souza, Moews, Dole, Castellano, Merlin, {et~al.}}]{boucaud2020photometry}
Boucaud, A., Huertas-Company, M., Heneka, C., {et~al.} 2020, Monthly Notices of the Royal Astronomical Society, 491, 2481

\bibitem[{Breckinridge {et~al.}(2015)Breckinridge, Lam, \& Chipman}]{breckinridge2015polarization}
Breckinridge, J.~B., Lam, W. S.~T., \& Chipman, R.~A. 2015, Publications of the Astronomical Society of the Pacific, 127, 445

\bibitem[{Buncher {et~al.}(2021)Buncher, Sharma, \& Carrasco~Kind}]{buncher2021survey2survey}
Buncher, B., Sharma, A.~N., \& Carrasco~Kind, M. 2021, Monthly Notices of the Royal Astronomical Society, 503, 777

\bibitem[{Cai {et~al.}(2019)Cai, Zeng, Yong, Cao, \& Zhang}]{cai2019toward}
Cai, J., Zeng, H., Yong, H., Cao, Z., \& Zhang, L. 2019, in Proceedings of the IEEE/CVF international conference on computer vision, 3086--3095

\bibitem[{Cantale {et~al.}(2016)Cantale, Courbin, Tewes, Jablonka, \& Meylan}]{cantale2016firedec}
Cantale, N., Courbin, F., Tewes, M., Jablonka, P., \& Meylan, G. 2016, Astronomy \& Astrophysics, 589, A81

\bibitem[{Chao {et~al.}(2019)Chao, Chang, Wang, Cheng, Deng, \& Duan}]{chao2019high}
Chao, W., Chang, L., Wang, X., {et~al.} 2019, in 2019 IEEE International Conference on Image Processing (ICIP), IEEE, 4699--4703

\bibitem[{Chen(2021)}]{chen2021challenges}
Chen, H. 2021in , IOP Publishing, 012066

\bibitem[{Chen \& Hays(2018)}]{chen2018sketchygan}
Chen, W., \& Hays, J. 2018, in Proceedings of the IEEE conference on computer vision and pattern recognition, 9416--9425

\bibitem[{Conselice(2003)}]{conselice2003relationship}
Conselice, C.~J. 2003, The Astrophysical Journal Supplement Series, 147, 1

\bibitem[{Conselice(2014)}]{conselice2014evolution}
---. 2014, Annual Review of Astronomy and Astrophysics, 52, 291

\bibitem[{Courbin {et~al.}(1999)Courbin, Magain, Sohy, Lidman, \& Meylan}]{courbin1999deconvolving}
Courbin, F., Magain, P., Sohy, S., Lidman, C., \& Meylan, G. 1999, Messenger, 97

\bibitem[{Dawson {et~al.}(2012)Dawson, Schlegel, Ahn, Anderson, Aubourg, Bailey, Barkhouser, Bautista, Beifiori, Berlind, {et~al.}}]{dawson2012baryon}
Dawson, K.~S., Schlegel, D.~J., Ahn, C.~P., {et~al.} 2012, The Astronomical Journal, 145, 10

\bibitem[{Demir \& Unal(2018)}]{demir2018patch}
Demir, U., \& Unal, G. 2018, arXiv preprint arXiv:1803.07422

\bibitem[{DePoy {et~al.}(2008)DePoy, Abbott, Annis, Antonik, Barcel{\'o}, Bernstein, Bigelow, Brooks, Buckley-Geer, Campa, {et~al.}}]{depoy2008dark}
DePoy, D., Abbott, T., Annis, J., {et~al.} 2008, in Ground-based and Airborne Instrumentation for Astronomy II, Vol. 7014, SPIE, 190--198

\bibitem[{Dey {et~al.}(2019)Dey, Schlegel, Lang, Blum, Burleigh, Fan, Findlay, Finkbeiner, Herrera, Juneau, {et~al.}}]{dey2019overview}
Dey, A., Schlegel, D.~J., Lang, D., {et~al.} 2019, The Astronomical Journal, 157, 168

\bibitem[{Dou {et~al.}(2022)Dou, Xu, Ren, Zhao, \& Zhang}]{dou2022super}
Dou, F., Xu, L., Ren, Z., Zhao, D., \& Zhang, X. 2022, Research in Astronomy and Astrophysics, 22, 085018

\bibitem[{Driver {et~al.}(2011)Driver, Hill, Kelvin, Robotham, Liske, Norberg, Baldry, Bamford, Hopkins, Loveday, {et~al.}}]{driver2011galaxy}
Driver, S.~P., Hill, D.~T., Kelvin, L.~S., {et~al.} 2011, Monthly Notices of the Royal Astronomical Society, 413, 971

\bibitem[{Early {et~al.}(2004)Early, Hyde, \& Baron}]{early2004twenty}
Early, J.~T., Hyde, R., \& Baron, R.~L. 2004, in UV/Optical/IR Space Telescopes: Innovative Technologies and Concepts, Vol. 5166, SPIE, 148--156

\bibitem[{Flaugher {et~al.}(2015)Flaugher, Diehl, Honscheid, Abbott, Alvarez, Angstadt, Annis, Antonik, Ballester, Beaufore, {et~al.}}]{flaugher2015dark}
Flaugher, B., Diehl, H., Honscheid, K., {et~al.} 2015, The Astronomical Journal, 150, 150

\bibitem[{Glindemann {et~al.}(2000)Glindemann, Hippler, Berkefeld, \& Hackenberg}]{glindemann2000adaptive}
Glindemann, A., Hippler, S., Berkefeld, T., \& Hackenberg, W. 2000, Experimental Astronomy, 10, 5

\bibitem[{Gulrajani {et~al.}(2017)Gulrajani, Ahmed, Arjovsky, Dumoulin, \& Courville}]{gulrajani2017improved}
Gulrajani, I., Ahmed, F., Arjovsky, M., Dumoulin, V., \& Courville, A.~C. 2017, Advances in neural information processing systems, 30

\bibitem[{Gunn {et~al.}(2006)Gunn, Siegmund, Mannery, Owen, Hull, Leger, Carey, Knapp, York, Boroski, {et~al.}}]{gunn20062}
Gunn, J.~E., Siegmund, W.~A., Mannery, E.~J., {et~al.} 2006, The Astronomical Journal, 131, 2332

\bibitem[{Henry {et~al.}(2021)Henry, Natalie, \& Madsen}]{henry2021pix2pix}
Henry, J., Natalie, T., \& Madsen, D. 2021, Research Gate Publication, 1

\bibitem[{Heusel {et~al.}(2017)Heusel, Ramsauer, Unterthiner, Nessler, \& Hochreiter}]{heusel2017gans}
Heusel, M., Ramsauer, H., Unterthiner, T., Nessler, B., \& Hochreiter, S. 2017, Advances in neural information processing systems, 30

\bibitem[{Hickson(2014)}]{hickson2014atmospheric}
Hickson, P. 2014, The Astronomy and Astrophysics Review, 22, 1

\bibitem[{Hippler(2019)}]{hippler2019adaptive}
Hippler, S. 2019, Journal of Astronomical Instrumentation, 8, 1950001

\bibitem[{Isola {et~al.}(2017)Isola, Zhu, Zhou, \& Efros}]{isola2017image}
Isola, P., Zhu, J.-Y., Zhou, T., \& Efros, A.~A. 2017, in Proceedings of the IEEE conference on computer vision and pattern recognition, 1125--1134

\bibitem[{Jia {et~al.}(2021)Jia, Ning, Sun, Yang, \& Cai}]{jia2021data}
Jia, P., Ning, R., Sun, R., Yang, X., \& Cai, D. 2021, Monthly Notices of the Royal Astronomical Society, 501, 291

\bibitem[{Kinakh {et~al.}(2024)Kinakh, Belousov, Qu{\'e}tant, Drozdova, Holotyak, Schaerer, \& Voloshynovskiy}]{kinakh2024hubble}
Kinakh, V., Belousov, Y., Qu{\'e}tant, G., {et~al.} 2024, Sensors, 24, 1151

\bibitem[{Kingma(2014)}]{kingma2014adam}
Kingma, D.~P. 2014, arXiv preprint arXiv:1412.6980

\bibitem[{Kitchin(2020)}]{kitchin2020astrophysical}
Kitchin, C.~R. 2020, Astrophysical techniques (CRC press)

\bibitem[{Kumar {et~al.}(2024)Kumar, Bhadula, Al-Farouni, Varshney, Sharma, \& Saraswat}]{kumar2024artistic}
Kumar, D.~P., Bhadula, S., Al-Farouni, M., {et~al.} 2024, in 2024 International Conference on Communication, Computer Sciences and Engineering (IC3SE), IEEE, 1351--1357

\bibitem[{Le~F{\`e}vre {et~al.}(2005)Le~F{\`e}vre, Vettolani, Garilli, Tresse, Bottini, Le~Brun, Maccagni, Picat, Scaramella, Scodeggio, {et~al.}}]{le2005vimos}
Le~F{\`e}vre, O., Vettolani, G., Garilli, B., {et~al.} 2005, Astronomy \& Astrophysics, 439, 845

\bibitem[{Ledig {et~al.}(2017)Ledig, Theis, Husz{\'a}r, Caballero, Cunningham, Acosta, Aitken, Tejani, Totz, Wang, {et~al.}}]{ledig2017photo}
Ledig, C., Theis, L., Husz{\'a}r, F., {et~al.} 2017, in Proceedings of the IEEE conference on computer vision and pattern recognition, 4681--4690

\bibitem[{Lesser(2015)}]{lesser2015summary}
Lesser, M. 2015, Publications of the Astronomical Society of the Pacific, 127, 1097

\bibitem[{Li {et~al.}(2022)Li, Liu, \& Deng}]{li2022self}
Li, W., Liu, Z., \& Deng, H. 2022, in 2022 IEEE 8th International Conference on Computer and Communications (ICCC), IEEE, 1977--1981

\bibitem[{Li {et~al.}(2018)Li, Peng, Bhanu, Zhang, \& He}]{li2018super}
Li, Z., Peng, Q., Bhanu, B., Zhang, Q., \& He, H. 2018, Astrophysics and Space Science, 363, 1

\bibitem[{Lim {et~al.}(2017)Lim, Son, Kim, Nah, \& Mu~Lee}]{lim2017enhanced}
Lim, B., Son, S., Kim, H., Nah, S., \& Mu~Lee, K. 2017, in Proceedings of the IEEE conference on computer vision and pattern recognition workshops, 136--144

\bibitem[{Lin {et~al.}(2021)Lin, Fouchez, \& Pasquet}]{lin2021galaxy}
Lin, Q., Fouchez, D., \& Pasquet, J. 2021, in 2020 25th International Conference on Pattern Recognition (ICPR), IEEE, 5634--5641

\bibitem[{Lintott {et~al.}(2011)Lintott, Schawinski, Bamford, Slosar, Land, Thomas, Edmondson, Masters, Nichol, Raddick, {et~al.}}]{lintott2011galaxy}
Lintott, C., Schawinski, K., Bamford, S., {et~al.} 2011, Monthly Notices of the Royal Astronomical Society, 410, 166

\bibitem[{Lintott {et~al.}(2008)Lintott, Schawinski, Slosar, Land, Bamford, Thomas, Raddick, Nichol, Szalay, Andreescu, {et~al.}}]{lintott2008galaxy}
Lintott, C.~J., Schawinski, K., Slosar, A., {et~al.} 2008, Monthly Notices of the Royal Astronomical Society, 389, 1179

\bibitem[{Liu {et~al.}(2020)Liu, Song, Zhu, \& Elgammal}]{liu2020sketch}
Liu, B., Song, K., Zhu, Y., \& Elgammal, A. 2020, in Proceedings of the Asian Conference on Computer Vision

\bibitem[{Liu \& Xu(2022)}]{liu2022night}
Liu, C., \& Xu, B. 2022, Computer-Aided Civil and Infrastructure Engineering, 37, 1737

\bibitem[{Long {et~al.}(2021)Long, Soubo, Cong, Weiping, \& Tong}]{long2021learning}
Long, M., Soubo, Y., Cong, S., Weiping, N., \& Tong, L. 2021, Monthly Notices of the Royal Astronomical Society, 504, 1077

\bibitem[{Ma {et~al.}(2014)Ma, Shang, Hu, Liu, Wang, \& Wei}]{ma2014new}
Ma, B., Shang, Z., Hu, Y., {et~al.} 2014, in High energy, optical, and infrared detectors for astronomy vi, Vol. 9154, SPIE, 593--600

\bibitem[{Madau \& Dickinson(2014)}]{madau2014cosmic}
Madau, P., \& Dickinson, M. 2014, Annual Review of Astronomy and Astrophysics, 52, 415

\bibitem[{Magain {et~al.}(2007)Magain, Courbin, Gillon, Sohy, Letawe, Chantry, \& Letawe}]{magain2007deconvolution}
Magain, P., Courbin, F., Gillon, M., {et~al.} 2007, Astronomy \& Astrophysics, 461, 373

\bibitem[{Magain {et~al.}(1998)Magain, Courbin, \& Sohy}]{magain1998deconvolution}
Magain, P., Courbin, F., \& Sohy, S. 1998, The Astrophysical Journal, 494, 472

\bibitem[{Massey {et~al.}(2010)Massey, Kitching, \& Richard}]{massey2010dark}
Massey, R., Kitching, T., \& Richard, J. 2010, Reports on Progress in Physics, 73, 086901

\bibitem[{Miao {et~al.}(2024)Miao, Tu, Jiang, Li, \& Qiu}]{miao2024astrosr}
Miao, J., Tu, L., Jiang, B., Li, X., \& Qiu, B. 2024, The Astrophysical Journal Supplement Series, 274, 7

\bibitem[{Miyazaki {et~al.}(2012)}]{miyazaki2012mclean}
Miyazaki, S., {et~al.} 2012, in Proc. SPIE Conf. Ser, Vol. 8446, 327

\bibitem[{Miyazaki {et~al.}(2018)Miyazaki, Komiyama, Kawanomoto, Doi, Furusawa, Hamana, Hayashi, Ikeda, Kamata, Karoji, {et~al.}}]{miyazaki2018hyper}
Miyazaki, S., Komiyama, Y., Kawanomoto, S., {et~al.} 2018, Publications of the Astronomical Society of Japan, 70, S1

\bibitem[{Morales {et~al.}(2018)Morales, Mart{\'\i}nez-Delgado, Grebel, Cooper, Javanmardi, \& Miskolczi}]{morales2018systematic}
Morales, G., Mart{\'\i}nez-Delgado, D., Grebel, E.~K., {et~al.} 2018, Astronomy \& Astrophysics, 614, A143

\bibitem[{Nyquist(1928)}]{nyquist1928certain}
Nyquist, H. 1928, Transactions of the American Institute of Electrical Engineers, 47, 617

\bibitem[{Paez \& Strojnik(2001)}]{paez2001telescopes}
Paez, G., \& Strojnik, M. 2001, in Handbook of Optical Engineering (CRC Press), 225--280

\bibitem[{Popowicz {et~al.}(2016)Popowicz, Kurek, Blachowicz, Orlov, \& Smolka}]{popowicz2016efficiency}
Popowicz, A., Kurek, A., Blachowicz, T., Orlov, V., \& Smolka, B. 2016, Monthly Notices of the Royal Astronomical Society, 463, 2172

\bibitem[{Reiman \& G{\"o}hre(2019)}]{reiman2019deblending}
Reiman, D.~M., \& G{\"o}hre, B.~E. 2019, Monthly Notices of the Royal Astronomical Society, 485, 2617

\bibitem[{Roggemann {et~al.}(1997)Roggemann, Welsh, \& Fugate}]{roggemann1997improving}
Roggemann, M.~C., Welsh, B.~M., \& Fugate, R.~Q. 1997, Reviews of Modern Physics, 69, 437

\bibitem[{Ronneberger {et~al.}(2015)Ronneberger, Fischer, \& Brox}]{ronneberger2015u}
Ronneberger, O., Fischer, P., \& Brox, T. 2015, in Medical image computing and computer-assisted intervention--MICCAI 2015: 18th international conference, Munich, Germany, October 5-9, 2015, proceedings, part III 18, Springer, 234--241

\bibitem[{Schawinski {et~al.}(2017)Schawinski, Zhang, Zhang, Fowler, \& Santhanam}]{schawinski2017generative}
Schawinski, K., Zhang, C., Zhang, H., Fowler, L., \& Santhanam, G.~K. 2017, Monthly Notices of the Royal Astronomical Society: Letters, 467, L110

\bibitem[{Shannon(1949)}]{shannon1949communication}
Shannon, C.~E. 1949, Proceedings of the IRE, 37, 10

\bibitem[{Sheikh {et~al.}(2004)}]{sheikh2004wang}
Sheikh, S., {et~al.} 2004, Image quality assessment: from error visibility to structural similarity, IEEE Trans. Image Process, 13, 600

\bibitem[{Shibuya {et~al.}(2024)Shibuya, Ito, Asai, Kirihara, Fujimoto, Toba, Miura, Umayahara, Iwadate, Ali, {et~al.}}]{shibuya2024galaxy}
Shibuya, T., Ito, Y., Asai, K., {et~al.} 2024, arXiv preprint arXiv:2403.06729

\bibitem[{Sreejith {et~al.}(2024)Sreejith, Slosar, \& Wang}]{sreejith2024point}
Sreejith, S., Slosar, A., \& Wang, H. 2024, Physical Review D, 110, 103030

\bibitem[{Starck {et~al.}(2002)Starck, Pantin, \& Murtagh}]{starck2002deconvolution}
Starck, J.-L., Pantin, E., \& Murtagh, F. 2002, Publications of the Astronomical Society of the Pacific, 114, 1051

\bibitem[{Sweere {et~al.}(2022)Sweere, Valtchanov, Lieu, Vojtekova, Verdugo, Santos-Lleo, Pacaud, Briassouli, \& C{\'a}mpora~P{\'e}rez}]{sweere2022deep}
Sweere, S.~F., Valtchanov, I., Lieu, M., {et~al.} 2022, Monthly Notices of the Royal Astronomical Society, 517, 4054

\bibitem[{Szegedy {et~al.}(2015)Szegedy, Liu, Jia, Sermanet, Reed, Anguelov, Erhan, Vanhoucke, \& Rabinovich}]{szegedy2015going}
Szegedy, C., Liu, W., Jia, Y., {et~al.} 2015, in Proceedings of the IEEE conference on computer vision and pattern recognition, 1--9

\bibitem[{Thanh-Tung \& Tran(2020)}]{thanh2020catastrophic}
Thanh-Tung, H., \& Tran, T. 2020, in 2020 international joint conference on neural networks (ijcnn), IEEE, 1--10

\bibitem[{Tirel {et~al.}(2024)Tirel, Ali, \& Hashim}]{tirel2024novel}
Tirel, L., Ali, A.~M., \& Hashim, H.~A. 2024, Systems and Soft Computing, 6, 200122

\bibitem[{Vojtekova {et~al.}(2021)Vojtekova, Lieu, Valtchanov, Altieri, Old, Chen, \& Hroch}]{vojtekova2021learning}
Vojtekova, A., Lieu, M., Valtchanov, I., {et~al.} 2021, Monthly Notices of the Royal Astronomical Society, 503, 3204

\bibitem[{Walmsley {et~al.}(2022)Walmsley, Lintott, G{\'e}ron, Kruk, Krawczyk, Willett, Bamford, Kelvin, Fortson, Gal, {et~al.}}]{walmsley2022galaxy}
Walmsley, M., Lintott, C., G{\'e}ron, T., {et~al.} 2022, Monthly Notices of the Royal Astronomical Society, 509, 3966

\bibitem[{Wang {et~al.}(2023)Wang, Sreejith, Lin, Ramachandra, Solsar, \& Yoo}]{wang2023neural}
Wang, H., Sreejith, S., Lin, Y., {et~al.} 2023, The Open Journal of Astrophysics, 6, 30

\bibitem[{Wang {et~al.}(2022)Wang, Sreejith, Slosar, Lin, \& Yoo}]{wang2022galaxy}
Wang, H., Sreejith, S., Slosar, A., Lin, Y., \& Yoo, S. 2022, Physical Review D, 106, 063023

\bibitem[{Willett {et~al.}(2013)Willett, Lintott, Bamford, Masters, Simmons, Casteels, Edmondson, Fortson, Kaviraj, Keel, {et~al.}}]{willett2013galaxy}
Willett, K.~W., Lintott, C.~J., Bamford, S.~P., {et~al.} 2013, Monthly Notices of the Royal Astronomical Society, 435, 2835

\bibitem[{Xia {et~al.}(2022)Xia, Hang, Tian, Yang, Liao, \& Zhou}]{xia2022efficient}
Xia, B., Hang, Y., Tian, Y., {et~al.} 2022in , 2759--2767

\bibitem[{Yang {et~al.}(2023)Yang, Chen, Tang, Deng, \& Wang}]{yang2023image}
Yang, Q., Chen, Z., Tang, R., Deng, X., \& Wang, J. 2023, The Astrophysical Journal Supplement Series, 265, 36

\bibitem[{York {et~al.}(2000)York, Adelman, Anderson~Jr, Anderson, Annis, Bahcall, Bakken, Barkhouser, Bastian, Berman, {et~al.}}]{york2000sloan}
York, D.~G., Adelman, J., Anderson~Jr, J.~E., {et~al.} 2000, The Astronomical Journal, 120, 1579

\bibitem[{Zhang {et~al.}(2018{\natexlab{a}})Zhang, Isola, Efros, Shechtman, \& Wang}]{zhang2018unreasonable}
Zhang, R., Isola, P., Efros, A.~A., Shechtman, E., \& Wang, O. 2018{\natexlab{a}}, in Proceedings of the IEEE conference on computer vision and pattern recognition, 586--595

\bibitem[{Zhang {et~al.}(2018{\natexlab{b}})Zhang, Li, Li, Wang, Zhong, \& Fu}]{zhang2018image}
Zhang, Y., Li, K., Li, K., {et~al.} 2018{\natexlab{b}}, in Proceedings of the European conference on computer vision (ECCV), 286--301

\bibitem[{Zibetti {et~al.}(2007)Zibetti, M{\'e}nard, Nestor, Quider, Rao, \& Turnshek}]{zibetti2007optical}
Zibetti, S., M{\'e}nard, B., Nestor, D.~B., {et~al.} 2007, The Astrophysical Journal, 658, 161

\end{thebibliography}
\bibliographystyle{aasjournal}

%% This command is needed to show the entire author+affiliation list when
%% the collaboration and author truncation commands are used.  It has to
%% go at the end of the manuscript.
%\allauthors

%% Include this line if you are using the \added, \replaced, \deleted
%% commands to see a summary list of all changes at the end of the article.
%\listofchanges

\end{document}